\documentclass[aps,reprint,twocolumn,pra,groupedaddress,floatfix]{revtex4-2}
\usepackage{amsmath,amssymb,amsfonts}
\usepackage{mathrsfs}
\usepackage{stackengine}
\usepackage{soul}
\usepackage{graphicx}
\usepackage{color}
\usepackage[table]{xcolor}

\graphicspath{{./Figsbroken/}}
\begin{document}
	\title{Unique multistable states in periodic structures with saturable nonlinearity. II. Broken $\mathcal{PT}$-symmetric regime }
	\author{S. Vignesh Raja}
	\author{A. Govindarajan}
	\email[Corresponding author: ]{govin.nld@gmail.com}
	\author{M. Lakshmanan}
	\affiliation{$^{*}$Department of Physics, Pondicherry University, Puducherry, 605014, India}
	
	\affiliation{$^{\dagger,\mathsection}$Department of Nonlinear Dynamics, School of Physics, Bharathidasan University, Tiruchirappalli - 620 024, India}
	\begin{abstract}
		 	 In this work, we observe that the $\mathcal{PT}$-symmetric fiber Bragg gratings (PTFBGs) with saturable nonlinearity (SNL) exhibit ramp-like, mixed, optical multistability (OM) in the broken regime.  The interplay between nonlinearity and detuning parameter plays a central role in transforming the characteristics of the hysteresis curves and facilitates the realization of different OM curves. Also, it plays a crucial role in reducing the switch-up and down intensities of various stable branches of an OM curve. In a mixed OM curve, either the ramp-like hysteresis curves or S-like hysteresis curves can appear predominantly  depending on the magnitude of the detuning parameter. An increase in the device length or nonlinearity increases the number of stable states for fixed values of input intensity. Under a reversal in the direction of light incidence, the ramp-like OM and mixed OM curves assume an unusual vortex-like envelope at lower intensities.  Numerical simulations reveal that the switch-up and down intensities of different stable branches of a ramp-like OM and mixed OM curves drift towards the higher and lower intensity sides, respectively (opposite direction). The drift is severe to the extent that an intermediate hysteresis curve features switch-down action at near-zero switching intensities. Also, the input intensities required to realize ramp-like, and mixed OM curves reduce dramatically under a reversal in the direction of light incidence. 
	\end{abstract}
	\maketitle
	\section{Introduction}
	
	Non-Hermitian physicists proposed a modern approach for dealing with losses by engineering a photonic platform that allows tailored light propagation via an interplay between the refractive index (RI) and balanced gain-loss levels of the medium \cite{el2007theory,regensburger2012parity,feng2017non,longhi2010pt}. These versatile devices are known as parity and time ($\mathcal{PT}$)- symmetric photonic systems, and they exhibit unusual optical dynamics such as nonreciprocal, unidirectional (reflectionless) wave transport and lasing behavior \cite{kottos2010optical}.  In such devices, the gain and loss regions are periodically placed in the neighborhood of each other to satisfy the $\mathcal{PT}$-symmetric RI condition, $n(z)=n^*(-z)$ \cite{ruter2010observation,govindarajan2018tailoring}. Investigating the tailored band structure, energy flow, and the resulting optical dynamics of $\mathcal{PT}$-symmetric systems are some hot topics of interest in modern-day research (see \cite{ozdemir2019parity,el2018non} and references therein). Among the wide range of $\mathcal{PT}$-symmetric photonic configurations, PTFBGs are one of the most fertile sources for testing various $\mathcal{PT}$-symmetric concepts \cite{phang2013ultrafast,phang2014impact,phang2015versatile}. 
	
	Before we look into the structure and operation of a PTFBG, let us recall the fundamental concepts related to the passive FBG. The periodic RI variations of a FBG lead to the reflection of a part of the incoming optical field (stopband)  \cite{erdogan1997fiber}. A background material with a constant RI ($n_o$) encloses the FBG of length $L = N \Lambda$, where $N$ and $\Lambda$ stand for the number of grating sections and length of one unit cell (also known as one grating period), respectively \cite{hill1997fiber}. A passive FBG works based on the Bragg reflection condition $\lambda_b = 2 n_0 \Lambda$, where $\lambda_b$  stands for the Bragg wavelength \cite{erdogan1997fiber}.  A passive grating offers endless possibilities to tailor its spectral characteristics according to the application \cite{erdogan1997fiber,hill1997fiber}. The research field of FBG is still flourishing with a considerable research interest, and new grating structures are evolving continuously with time.  With the introduction of $\mathcal{PT}$-symmetric notions, the research field of FBG has reached a milestone in its progress.
	
	Poladian \emph{et al.} and Kulishov \emph{et al.} have respectively proposed the original mathematical formulation and theoretical design of the FBGs with gain and loss, which are now known as PTFBGs \cite{poladian1996resonance,kulishov2005nonreciprocal}. The theory of $\mathcal{PT}$- symmetry suggests that the RI profile [$n(z)$] should be a complex quantity to satisfy the $\mathcal{PT}$-symmetric condition. The imaginary part of the RI profile signifies the gain and loss levels ($g = \pi n_{1I}/\lambda$) fed to the system and is a function of the operating wavelength ($\lambda$). The modulation of the imaginary part of the RI [$n_{1I}$ sin(2 $\pi z/\Lambda$)] is an odd function of the spatial coordinate ($z$). Recall that the coupling strength ($\kappa = \pi n_{1R}/\lambda$) is a function of the real part of the RI ($n_{1R}$) and the operating wavelength ($\lambda$).  The corresponding modulation term [$n_{1R}$ cos(2 $\pi z/\Lambda$)] is an even function of $z$ \cite{miri2012bragg,sarma2014modulation}. A background material with a uniform RI $n_0$ encloses the gain and loss segments of a PTFBG. The structural difference between these two configurations lies in the architecture of a unit cell \cite{phang2013ultrafast,phang2014impact}. A conventional FBG features alternate regions of modulated and unmodulated RIs, whereas a PTFBG features alternate gain and loss regions in one grating period \cite{raja2020tailoring,raja2020phase,raja2021n,phang2013ultrafast,phang2014impact,phang2015versatile}. The notion of $\mathcal{PT}$-symmetry enables the study of the light dynamics of a particular configuration in three different operating regimes, namely unbroken, broken regimes, and the unitary transmission point, which enables the same system to offer multi-functionalities against a variation in the operating conditions \cite{lin2011unidirectional,phang2013ultrafast,kottos2010optical,regensburger2012parity,el2007theory,ruter2010observation,feng2017non,raja2021n}.  
	
	In the past, researchers have investigated PTFBGs in the nonlinear regime on various themes, including the study of gap and Bragg soliton formation \cite{miri2012bragg,raja2019multifaceted}, nonlinear spectra \cite{raja2019multifaceted}, modulational instability \cite{sarma2014modulation}, and optical switching \cite{PhysRevA.100.053806,sudhakar2022inhomogeneous,sudhakar2022low,1555-6611-25-1-015102,komissarova2019pt}. Among these, optical switching serves as the central theme of the present work. The literature presents a wide range of options to optimize the power required for switching. Frequency detuning, an old-age technique proposed by Winful \emph{et al.,} stands out as one of the best tools to reduce the switching intensities in any nonlinear periodic structure even today \cite{winful1979theory}. The system's operation in the negative and positive detuning regimes in the presence of self-defocusing and focusing nonlinearities leads to low-power switching, respectively \cite{raja2019multifaceted,PhysRevA.100.053806}. Design optimization is another perspective from which researchers have studied the switching dynamics of periodic structures, which includes the discovery of low-power all-optical switching in conventional phase-shifted FBGs (PSFBGs), $\mathcal{PT}$-symmetric PSFBGs \cite{radic1994optical,ramakrishnan2019bistability}, conventional and $\mathcal{PT}$-symmetric chirped FBGs \cite{maywar1998effect, PhysRevA.100.053806}. The researchers have also investigated the switching dynamics of FBGs in the time-domain by manipulating signal parameters like pulse width and shape.  \cite{lee2003nonlinear}.
	
	 It is well-known that the nonlinearity in fibers originates from the intensity-dependent RI of the medium. For instance, a silica fiber supports only lower-order cubic nonlinearity \cite{1555-6611-25-1-015102}, whereas a chalcogenide glass gives rise to higher-order nonlinearities such as quintic and septimal ones \cite{chen2006measurement,harbold2002highly,raja2019multifaceted,PhysRevA.100.053806,sudhakar2022inhomogeneous,sudhakar2022low}.  Instead of having homogeneous nonlinear profiles, we can create structures with inhomogeneous nonlinearities by varying the dopant concentration as a function of spatial coordinates during fabrication \cite{sudhakar2022inhomogeneous}. In these works, the SNL does not play a significant role in altering the nonlinear response of the system. But in some cases, the inclusion of the SNL effect is necessary to characterize the nonlinear response of the glass fiber. For example, sulfide and heavy-metal oxide doped glasses exhibit SNL owing to their faster nonlinear response time, and slower thermal response \cite{acioli1988measurement,coutaz1991saturation,hall1989nonlinear,yao1985ultrafast}. The intensity at which SNL comes into effect varies from material to material \cite{abou2011impact}. In the case of $CdS_{x}Se_{1-x}$ semiconductor-doped glass fibers \cite{gatz1991soliton,olbright1986interferometric}, SNL comes into the picture at moderate intensities, whereas heavy-metal-doped oxide fibers \cite{kang1995femtosecond} require high intensity for saturation of the nonlinear response to occur.

 The studies on the SNL in periodic structures have primarily focused on the soliton formation, collision dynamics, and stability of solitons \cite{malomed2005coupled}. It is well-known that the phenomenon of OB goes hand in hand with the soliton formation in periodic structures \cite{raja2019multifaceted}. However, the literature seems to lack comprehensive studies on the OB (OM) phenomenon in FBGs with SNL. Recently, in a previous work, we have studied the nonlinear transmission characteristics of FBG and PTFBG with SNL in the conventional system and unbroken $\mathcal{PT}$- symmetric regime \cite{raja2022saturate}. We discovered that the hysteresis curves exhibited by a conventional FBG did not have typical S-shaped hysteresis curves. Instead, we have discovered that the input-output characteristics curves feature a ramp-like first stable state followed by hysteresis curves in which the output showed sharp variations against the input intensities (ramp-like OM curves) even in the conventional systems, which were not predicted in the literature before. Additionally, the width of the successive hysteresis curves gets broadened with an increase in the input intensities. The interplay between frequency detuning and $\mathcal{PT}$-symmetry becomes a requirement to retrieve the conventional S-shaped hysteresis curves. For smaller values of the detuning parameter, the system supports a new type of OM curve in which ramp-like stable states precede S-shaped hysteresis curves in the presence of $\mathcal{PT}$ symmetry.  Launching the input light from the rear end of the device resulted in a dramatic decrease in the switching at intensities to less than 0.001 in the unbroken $\mathcal{PT}$-symmetric regime, which is the lowest one ever reported in the literature to the best of our knowledge.

It would be interesting to know the effect of broken $\mathcal{PT}$-symmetry on the nonlinear transmission characteristics of FBG in the presence of SNL. Recent studies confirmed that the broken PTFBG supports OB (OM) curves with ramp-like first stable states under suitable manipulation of the system parameters \cite{raja2019multifaceted, PhysRevA.100.053806}.  Whether these states will continue to exist in the presence of SNL is a curious query, and Sec. \ref{Sec:3} addresses the question. The role of $\mathcal{PT}$-symmetry also paves the way to study the nonreciprocal switching dynamics in the presence of SNL, and Sec. \ref{Sec:4} deals with it. Section \ref{Sec:5} summarizes the crucial outcomes of the numerical studies.

\section{Theoretical framework}
\label{Sec:2}
The RI distribution [$n(z)$]  of a PTFBG that includes the SNL effect reads
\begin{gather}
	\nonumber n(z)=n_0+n_{1R} cos(2 \pi z/\Lambda)+in_{1I} sin(2 \pi z/\Lambda)\\-{n_2}f(|E|^2),
	\label{Eq:1}
\end{gather}

where $f(|E|^2)=\cfrac{1}{1+|E|^2}$ is the SNL function for periodic structures \cite{malomed2005coupled,merhasin2007solitons,yulin2008discrete,melvin2006radiationless,vicencio2006discrete,hadvzievski2004power}.  In Eq. (\ref{Eq:1}), $n_0$ and $n_2$ denote the RI of the core and nonlinear coefficient, respectively.  A detailed derivation of the coupled mode equations (CMEs) of the proposed system can be found in Ref. \cite{raja2022saturate}, and it reads 
\begin{gather}
	\frac{d u}{dz}=i\delta u+i \left(k+g\right)v-\cfrac{i S u}{(1+ |u|^{2}+|v|^{2})},
	\label{Eq:2}
\end{gather}
\begin{gather}
	-	\frac{d v}{dz}=i\delta v+i \left(k-g\right)u-\cfrac{i S v}{(1+ |u|^{2}+|v|^{2})}.
	\label{Eq:3}
\end{gather}
	These equations are valid for the left light incident condition. Under a reversal in the direction of light incidence, the term $\kappa + g$ in Eq. (\ref{Eq:2}) changes to $\kappa - g$. Also, the term $\kappa - g$ in Eq. (\ref{Eq:3}) is replaced by $\kappa + g$.
The presence of gain and loss terms distinguishes the final forms of CMEs in the present work from the governing model in Ref. \cite{malomed2005coupled,merhasin2007solitons}. The nonlinear coefficient of the material ($n_2$) gets transformed into a nonlinearity parameter $\left(S = \cfrac{2 \pi n_2}{\lambda}\right)$ while deriving the CMEs \cite{raja2019multifaceted,PhysRevA.100.053806,sarma2014modulation}. 

The system can be detuned on either side of the Bragg wavelength and is mathematically given by  $\delta=k-\pi/\Lambda=2\pi n_0\left(\cfrac{1}{\lambda}-\cfrac{1}{\lambda_b}\right)$. 

Using the implicit Runge-Kutta fourth-order method, the CMEs in (\ref{Eq:2}) and (\ref{Eq:3}) are integrated with the boundary conditions
\begin{gather}
	u(0)=u_0 \quad \text{and} \quad v(L)=0.
	\label{Eq:10}
\end{gather}
The input and output intensities read $P_0 = |u_0|^2$ and $P_1(L)=|u(L)|^2$, respectively. The device parameters used in the simulations are tabulated in Table \ref{tab1} and the reasons behind the selection of the values of each control parameter can be found in detail in Ref. \cite{raja2022saturate}.

\subsection{Need for separate investigation}
 This article covers the input-output characteristics of the PTFBG in the broken $\mathcal{PT}$-symmetric regime ($g > \kappa$). The decision to present the results pertaining to the input-output characteristics of the PTFBG with SNL operating in the broken $\mathcal{PT}$- symmetric regime separately, rather than combining them with the results obtained in the conventional and unbroken $\mathcal{PT}$- symmetric regimes, is justified by several crucial considerations. Firstly, in the traditional scenarios, the role of a control parameter on the OB and OM curves is typically uniform and straightforward to interpret, where the dynamics observed under the manipulation of a given system parameter tend to remain consistent across a wider range of variations in other control parameters. However, in the present system, this uniformity is not respected.

 To elloborate further, these system parameters, instead of acting independently, consistently work in conjunction with the SNL parameter, resulting in random variations in the OB and OM curves even for slight changes in their values. For instance, the detuning parameter, which conventionally reduces switching intensities across a wide range of nonlinearity variations \cite{raja2019multifaceted,PhysRevA.100.053806}, exhibits a different behavior in the present system in the unbroken $\mathcal{PT}$- symmetric regime \cite{raja2022saturate}.  Beyond its role in intensity reduction, it plays a pivotal role in transforming the nature of the hysteresis curve. This transformation varies from a ramp-like OB and OM curve at the synchronous wavelength, to mixed OM curves for smaller detuning parameter values, and finally, the retrieval of a conventional S-shaped curve becomes feasible only when the detuning parameter is sufficiently large. Even with the detuning parameter fixed in the case of a mixed OM curve, its behavior diverges for low and high intensities. This divergence leads to the generation of a new type of OB/OM curves that uniquely combine the features of well-known types of OM curves, namely, ramp-like and S-shaped ones.

This complexity necessitates a thorough investigation of the system dynamics within a narrow range of variations in other control parameters.  The division of results into two separate articles is driven by the need to present a detailed and focused analysis of the distinct dynamics observed in the broken $\mathcal{PT}$- symmetric regime, ensuring that each aspect of the system is thoroughly examined and comprehended.
 \subsection{Organization of the results}

  As we have already stated in the aforementioned section that the system parameters are highly sensitive with the changes in the value of SNL, we would like present this section exclusively as to how we have organized the simulations results in order to provide a clear picture on the upcoming numerical simulations. In Sec. \ref{Sec:3A}, we keep the value of the SNL parameter at $S = 0.5$ $W^{-1}/cm$ (for simplicity) and vary the detuning parameter to analyze its impact on the formation of S-shaped OB curves characterized by a gradual increase in the output intensities against input intensity tuning. Section \ref{Sec:3B} investigates the effect of increasing nonlinearity ($S = 1$ $W^{-1}/cm$) on the creation of ramp-like OB and OM curves characterized by a sharp or abrupt increase in the output intensities against input intensity tuning. Additionally, Sec. \ref{Sec:3C} explores the attribute of mixed OM curves (fusion of ramp-like and S-shaped OM) that feature sharp variations in the output for low intensities (similar to ramp-like OM) and gradual variations in the output for high intensities (similar to S-shaped OM) at $S = 2$ $W^{-1}/cm$. Examining the influence of device length on the number of stable states in the OM curves is covered in Sec. \ref{Sec:3D}. We chose the value of gain and loss parameter to be $g = 0.55$ $cm^{-1}$ exceeding the value of the coupling coefficient throughout the article unless specified. Section \ref{Sec:3E} highlights the detrimental impact of larger gain and loss parameters on the OB and OM curves.  Sections \ref{Sec:4A}, \ref{Sec:4B}, \ref{Sec:4C} and  \ref{Sec:4D} focus on the examination of low-power S-shaped OB, ramp-like OB, ramp-like OM and mixed OM curves curves, respectively, under the condition of right light incidence at $L = 20$ $cm$. To provide a high-level analysis of the problem, this section covers continuous variation plots of the switching intensities under two conditions. In the first scenario, the SNL parameter is fixed, and the detuning parameter is continuously varied. In the second scenario, the detuning parameter is fixed, and the SNL is varied continuously. Conducting this study is important to demonstrate that the numerical parameters chosen are not merely discrete values. Sections \ref{Sec:4E} -- \ref{Sec:4G} delve into distinct OM curve behaviors arising from the interplay between right light incidence conditions and larger device lengths, specifically focusing on the formation and variations in ramp-like and mixed OM curves with a vortex-like envelope.

 		\begin{table}[h!]
 		\caption{\centering{Values of device parameters used in the simulations} }
 		\begin{center}
 		
 			\begin{tabular}{c c c}
 				\hline
 				\hline
 				{Symbol}&{Device } & {Physical}  \\
 				{}&{parameter} & {values}  \\
 				\hline
 				
 				{$L$}&{device length}& {20 and 70 $cm$} \\

 				{$\kappa$}&{coupling coefficient}& {0.4 $cm^{-1}$}\\

 				{$g$}&{gain-loss }& {0.55 $cm$ $^{-1}$} \\
 				{}&{coefficient}&{(unless specified)}\\
 				
 				{$S$}&{nonlinearity coefficient}&{0.5, 1 and 2 $W^{-1}/cm$}\\

 				{$\lambda_b$}&{Bragg wavelength} & {1060 nm} \\
 				
 				{$\lambda_b$$-$$\lambda$}&{variations in operating} & {$\pm$ 0.02 nm} \\
 				{}&{ wavelength ($\lambda$) from $\lambda_b$}& {} \\

 				{$\delta$}&{detuning parameter}& {0 -- 2.5 $cm^{-1}$} \\
 				%

 				{$P_0$}&{input intensity }&{0 -- 1.5 $MW/cm^2$}\\
 				%
 				\hline\hline
 			\end{tabular}
 			\label{tab1}
 		\end{center}
 	\end{table}

\section{OB in the broken $\mathcal{PT}$-symmetric regime: Left incidence}\label{Sec:3}
 
\subsection{S-shaped OB curves}
\label{Sec:3A}
\begin{figure}[ht]
	\centering	\includegraphics[width=0.5\linewidth]{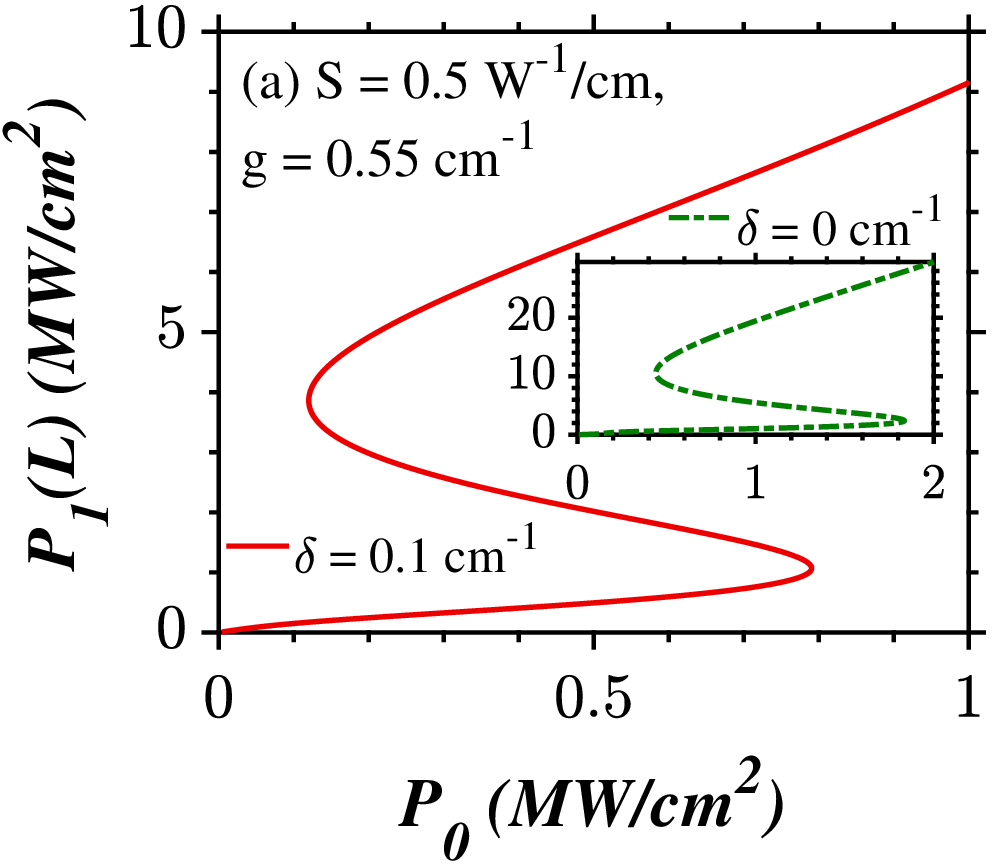}\includegraphics[width=0.5\linewidth]{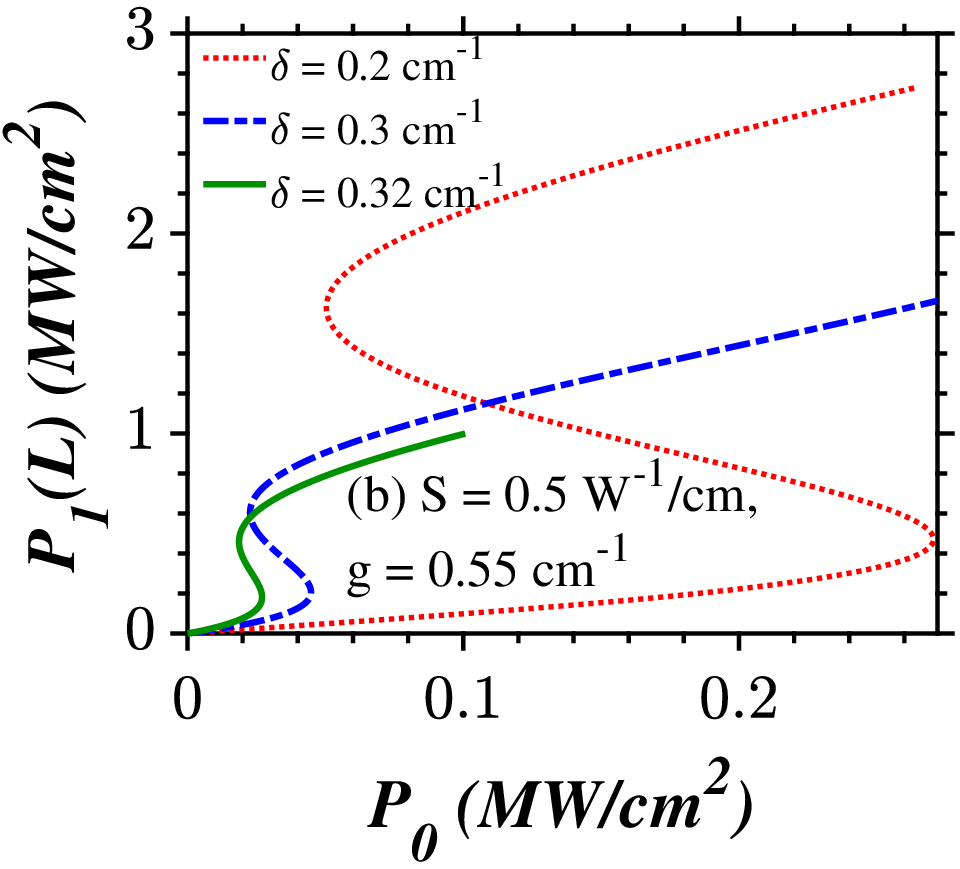}
	\caption{S-shaped OB curves in the input-output characteristics of a broken PTFBG ($g = 0.55$ $cm^{-1}$) with SNL at $L = 20$ cm and $S = 0.5$ $W^{-1}/cm$.  The light launching direction is left. The inset in (a) is simulated at the synchronous wavelength.}   
	\label{fig1}
\end{figure}

The system admits an S-shaped OB curve at $\delta = 0$ in the case of broken PTFBG with SNL, as shown in Figs. \ref{fig1}(a) and (b). As we tune the input intensity, the output intensity increases gradually in the first stable branch for a wide range of input intensities. When the input power exceeds a critical value known as switch-up intensity, we observe a jump in the output intensity from the first stable branch to the second. This mechanism is known as the switch-up mechanism. For any further increase in the input intensity, the output intensity continues to increase steadily along the second branch, and no further switching occurs with a rise in the input intensity. With a decrease in the input intensity, the output state does not jump back to the first stable branch at the switch-up intensity, and it remains in the second stable branch until the value of input intensity decreases below the switch-down intensity. The output state switches back to the first stable branch by the switch-down mechanism. The output is bistable for a given input intensity between the switch-up and down intensities. 

It should be recalled that OB refers to the system's ability to maintain only two stable states for specific input intensities, causing the output to vary between these two states. Consequently, the curves depicting this behavior are referred to as bistable curves. On the other hand, if the system demonstrates the capability to support more than two stable states for specific intensity values, resulting in the output varying among these multiple states, it is denoted as OM. The associated curves are labeled as multistable curves. The system does not favor OM formation with intensity tuning in Figs. \ref{fig1}(a) and (b) because the value of NL is too small to induce  multiple switching. 

When considering PTFBGs operating in the broken $\mathcal{PT}$- symmetric regime, the generation of S-shaped OB curves is indeed an uncommon occurrence in these systems, deviating from the expected behavior reported in the scientific literature \cite{raja2019multifaceted,PhysRevA.100.053806,sudhakar2022inhomogeneous,sudhakar2022low}. In our earlier publication focusing explicitly on OB/OM formation in the unbroken $\mathcal{PT}$- symmetric  regime, we have established that PTFBGs with SNL did not exhibit S-shaped OB and OM curves unless the system was operated at wavelengths far away from the Bragg wavelength via the frequency detuning \cite{raja2022saturate}. If this condition is not met, the system tends to favor the formation of OB and OM curves characterized by ramp-like first stable states, rather than S-shaped OB and OM curves. The inclusion of SNL guarantees that the PTFBG system produces ramp-like input-output characteristics in the unbroken $\mathcal{PT}$- symmetric regime, as seen in Ref. \cite{raja2022saturate}, and S-shaped OB/OM curves in the broken $\mathcal{PT}$- symmetric regime, as depicted in Fig. \ref{fig1}. This behavior stands in clear contrast to the input-output characteristics of PTFBGs without SNL which is not reported in any conventional systems before. 
\subsubsection{Impact of frequency detuning on the switching intensities }
\label{Sec:3A1}
Throughout the article, we study the OB (OM) phenomenon with positive values of the detuning parameter alone. An increase in the value of the detuning parameter reduces the switch-up and down intensities of the OB curve, as shown in Fig. \ref{fig1}(b). At this juncture, one may think that increasing the detuning parameter to a very large extent may result in ultralow power switching. Nevertheless, increasing the value of detuning parameter beyond a certain limit does not bring about this desired result; rather, it results in the disappearance of the OB behavior. Let the range of detuning parameters (detuning span) for which the system admits desirable S-shaped OB curves be $\delta_{span}^{S}$.  For S-shaped OB curves  generated at $S= 0.5$ $W^{-1}$/cm, the detuning span ranges from $\approx$ 0 to 0.32 $cm^{-1}$, as confirmed by the plots in Figs. \ref{fig1}(a) and (b).

The disappearance of OB curves for larger values of detuning parameter does not imply a complete elimination of the system's potential for further manipulating the switching dynamics as FBGs offers multiple avenues for altering the steering dynamics. However, it is essential to carefully adjust the other system parameters in the numerical simulations for optimal results, as detailed in the following sections. In such conditions, we are also keen to investigate whether it can reveal additional distinctive features not previously reported in the context of PTFBGs.

\subsection{Ramp-like OB curves}
\label{Sec:3B}

\begin{figure}
	\centering	\includegraphics[width=0.5\linewidth]{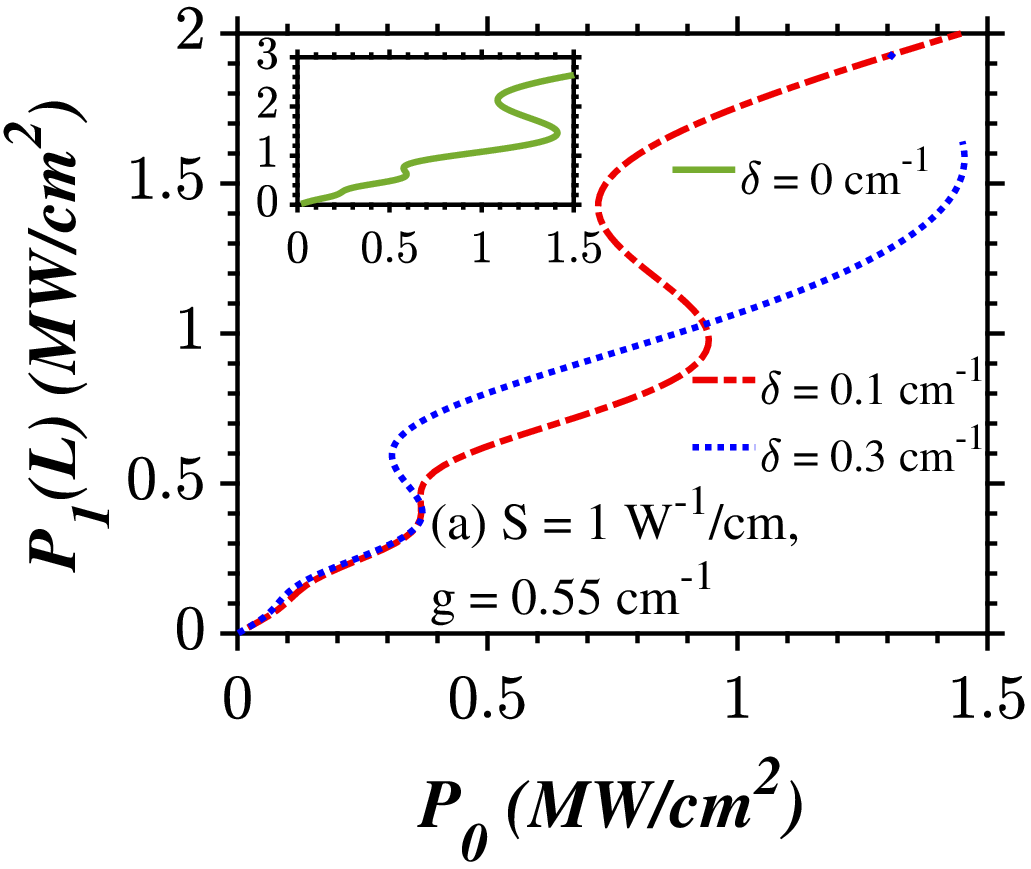}\includegraphics[width=0.5\linewidth]{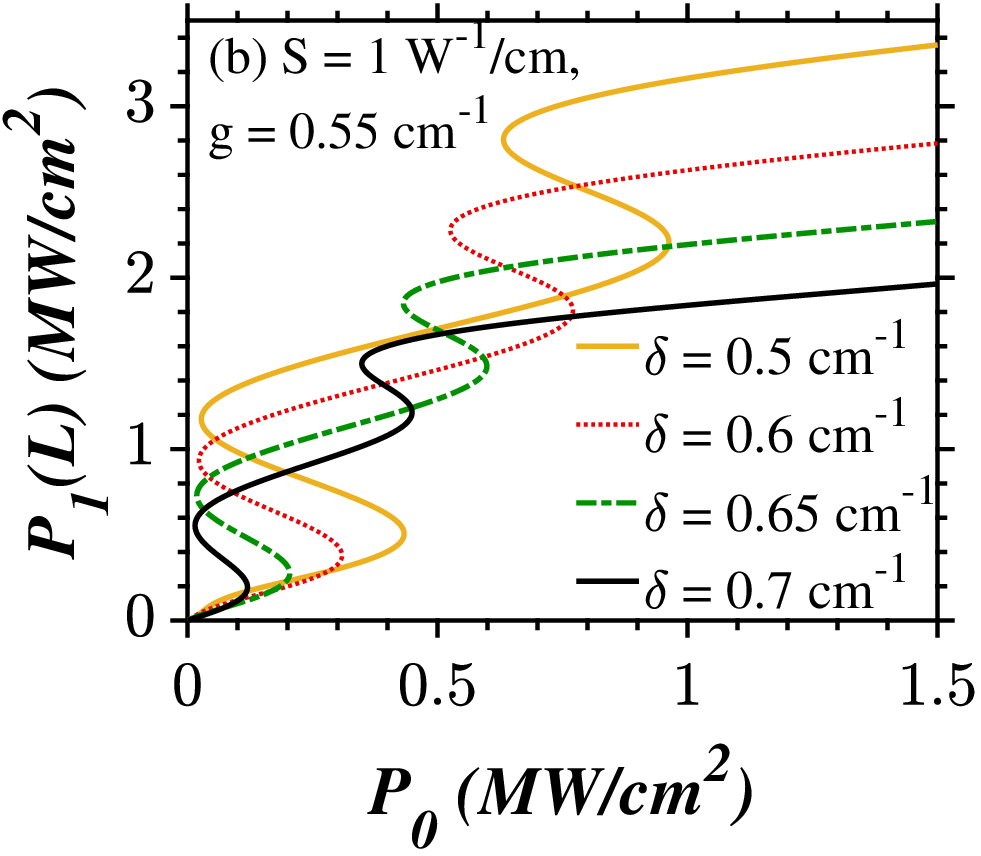}
	\caption{Input-output characteristics of a broken PTFBG ($g = 0.55$) with SNL at $L = 20$. The direction of light incidence is left. (a) Ramp-like OB curves and (b) ramp-like OM curves. The inset in (a) is simulated at the synchronous wavelength.}  
	\label{fig2}
\end{figure}

 The S-shaped OB curves in Fig. \ref{fig1} feature a gradual variation in the output intensities. In contrast, the plots in Fig. \ref{fig2}(a) features a ramp-like (sharp) increase in the output intensities along the first stable branch, and is named as ramp-like OB curve. Along these lines, ramp-like OM curves refer to hysteresis curves in which multiple stable branches appear on the top of a ramp-like first stable state. Within each branch, sharp and abrupt variations in the output intensities are observed against the input intensity. Moreover, as the input intensity is tuned, there is a noticeable widening of the successive hysteresis curves,   as shown in Fig. \ref{fig2}(b). An increase in the detuning parameter reduces the switch-up and down intensities of various stable branches of the ramp-like OB and OM curve in Figs. \ref{fig2}(a) and (b), respectively.  Let the range of detuning parameters for which the system admits ramp-like OB and OM curves be $\delta_{span}^{ramp}$ and it ranges from $\approx$ 0 to 1.5 $cm^{-1}$, as confirmed by the numerical simulations.

It is pertinent to note that these ramp-like OM curves also appear in the input-output characteristics of the broken PTFBG with SNL with $S = 2$ $W^{-1}/cm$ for $\delta < 1.2$ $cm^{-1}$, as shown in Fig. \ref{fig3}(a). Compared to Fig. \ref{fig2}(b), the number of stable states pertaining to the ramp-like OM curves in Fig. \ref{fig3}(a) is high, thanks to the increase in the value of nonlinearity parameter. 

It is important to recall that PTFBGs without SNL also feature ramp-like OB (OM) curves \cite{raja2019multifaceted,PhysRevA.100.053806}. The difference between the ramp-like OM curves exhibited by a PTFBG without SNL and with SNL lies in the characteristics of the hysteresis curves. The widths of the successive hysteresis curves decrease with an increase in the input intensity, as predicted in Refs. \cite{raja2019multifaceted, PhysRevA.100.053806}. On the other hand, the hysteresis width increases under similar conditions in the presence of SNL.  Therefore, the first and the $n^{th}$-order hysteresis curves feature the narrowest and broadest hysteresis widths, respectively, as shown in Figs. \ref{fig2}(b).  From these figures, we also observe that the range of detuning parameter within which the ramp-like OB/OM curves ($\delta_{span}^{ramp}$) occur is broader compared to the detuning span corresponding to the S-shaped OB curves ($\delta_{span}^{ramp}$), which indicates that the range of detuning parameter, in which favorable OB and OM curves appear, expands with an increase in the value of SNL.

\subsection{Mixed OM curves}
\label{Sec:3C}
\begin{figure}[t]
	\centering	\includegraphics[width=0.5\linewidth]{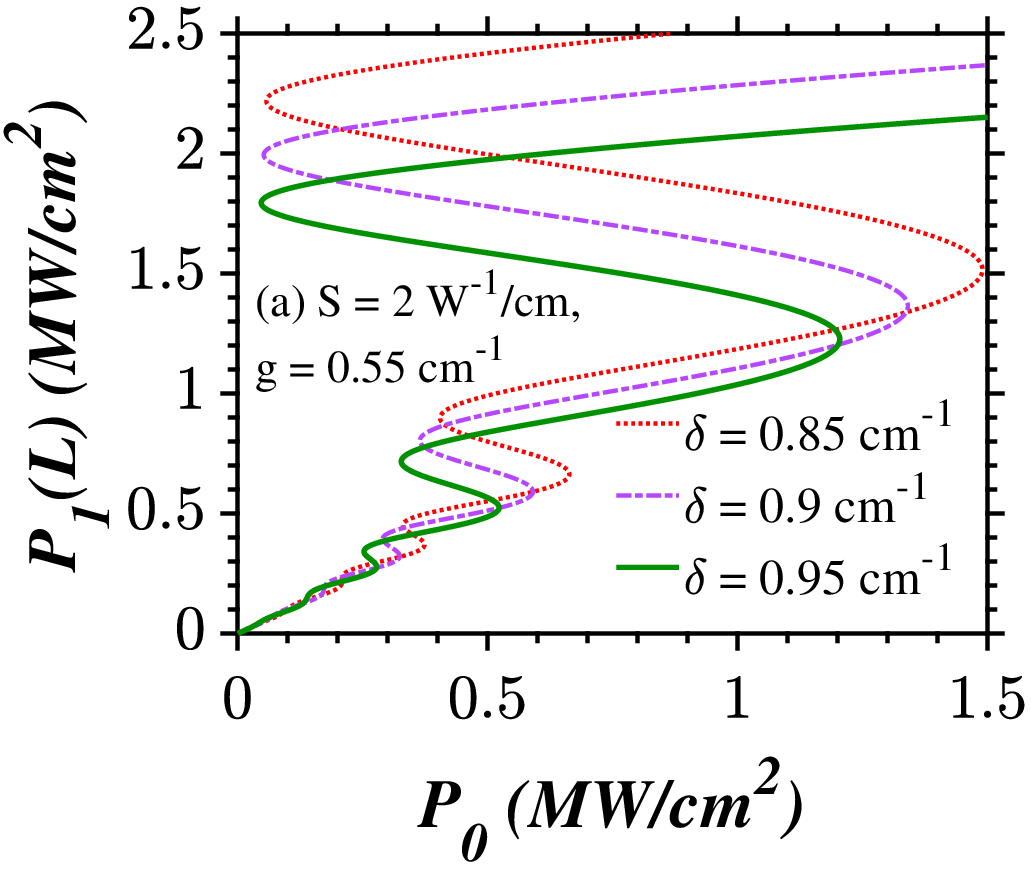}\includegraphics[width=0.5\linewidth]{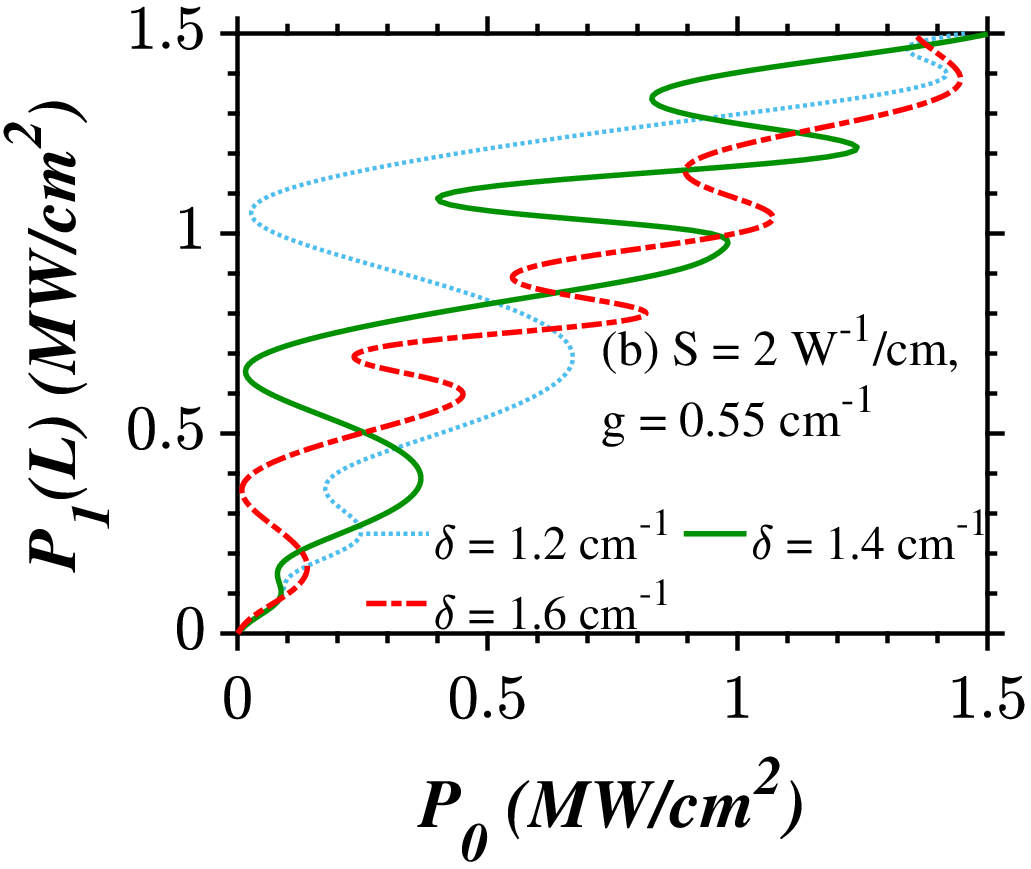}\\\includegraphics[width=0.5\linewidth]{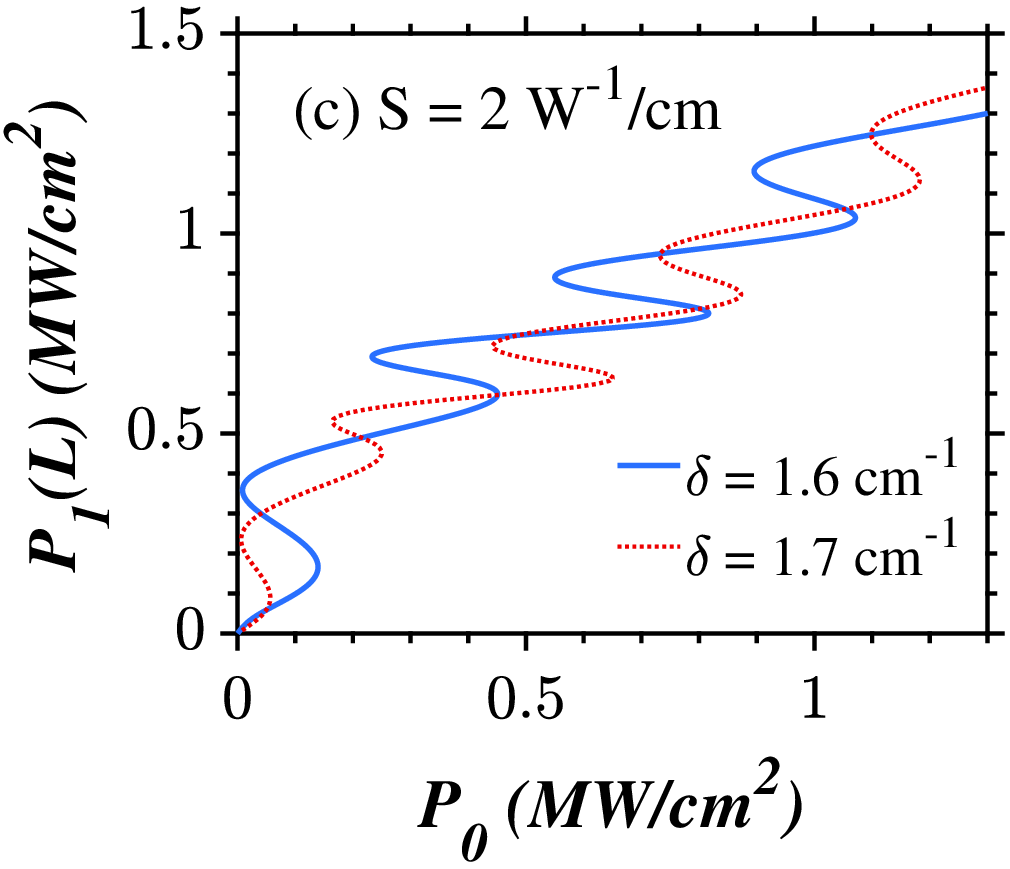}
	\caption{Frequency detuning induced (a) ramp-like (b) and (c) mixed OM curves in the input-output characteristics of a broken PTFBG with SNL at $L = 20$ $cm$ and $S = 2$ $W^{-1}/cm$. The direction of light incidence is left.} 
	\label{fig3}
\end{figure}

In Fig. \ref{fig3}(a), the hysteresis curve featured a sharp (ramp-like) increase in the output intensities against the variations in the input intensities. As the value of detuning parameter is increased beyond $\delta > 1.2$ $cm^{-1}$, the input-output characteristics of the system, depicted in Figs. \ref{fig3}(b), reveal mixed OM curves. This signifies a notable alteration in the nature of the hysteresis curves associated with changes in the detuning parameter value. Initially, the output intensities vary sharply against input intensities, a typical feature of ramp-like OB/OM curves. However, the same curve exhibits a gradual increase in output intensities for higher input values, a common characteristic found in S-shaped hysteresis curves. Since the same OM curve possess both the characteristics of ramp-like and S-shaped hysteresis curves, it is referred as the mixed OM curve. Let the range of detuning parameters for which the system admits the mixed OM curves be $\delta^{mix}_{span}$.   The switching intensities corresponding to the different stable branches get reduced considerably with an increase in the value of the detuning parameter, as shown in Fig. \ref{fig3}(c). Under the same condition, the number of S-shaped hysteresis curves in a mixed OM curve also increases, provided that the value of input intensity remains the same.

Thus, we can generate different OM curves like S-shaped, ramp-like, and mixed OM in the same system by tuning the values of saturable nonlinearity and detuning parameters and the results are summarized in Table \ref{tab2} for easier understanding the different types of OB/OM curves exhibited by the PTFBG in the broken $\mathcal{PT}$- symmetric regime.
 
\subsection{Role of device length on the number of stable states}
\label{Sec:3D}
\subsubsection{ Transformation of OB into ramp-like OM curves: $S < 1$ $W^{-1}/cm$}
\label{Sec:3E1}
\begin{figure}
	\centering	\includegraphics[width=0.5\linewidth]{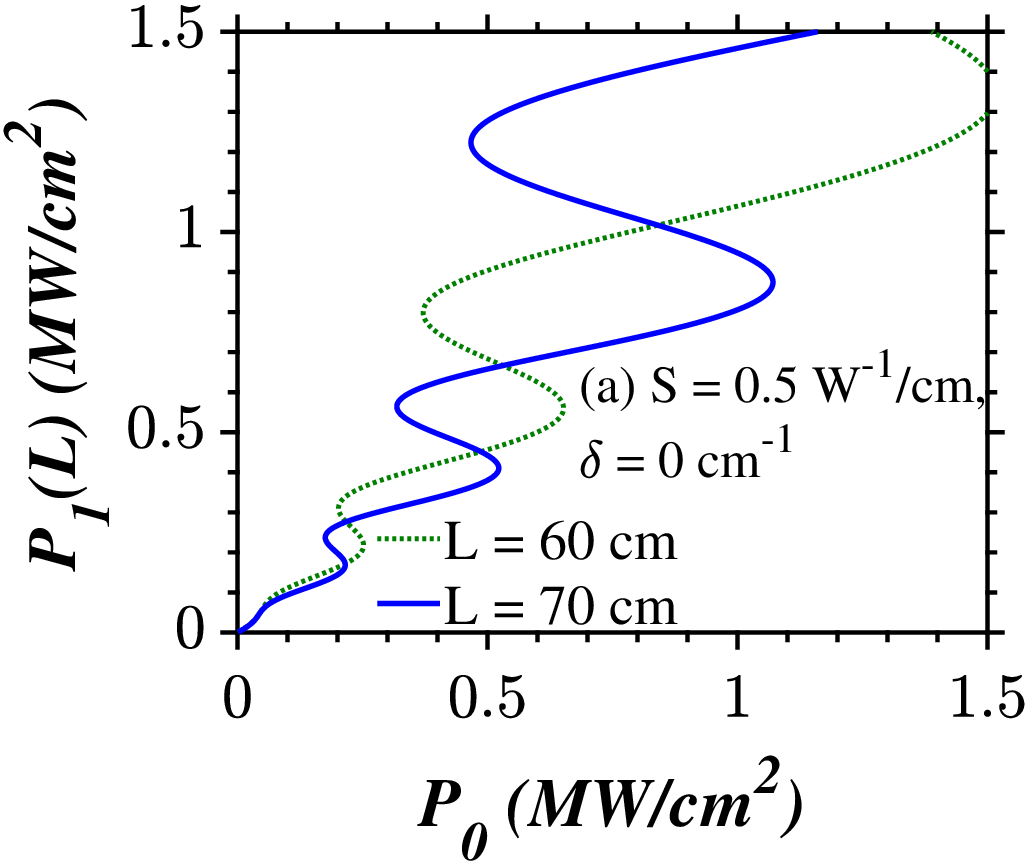}\includegraphics[width=0.5\linewidth]{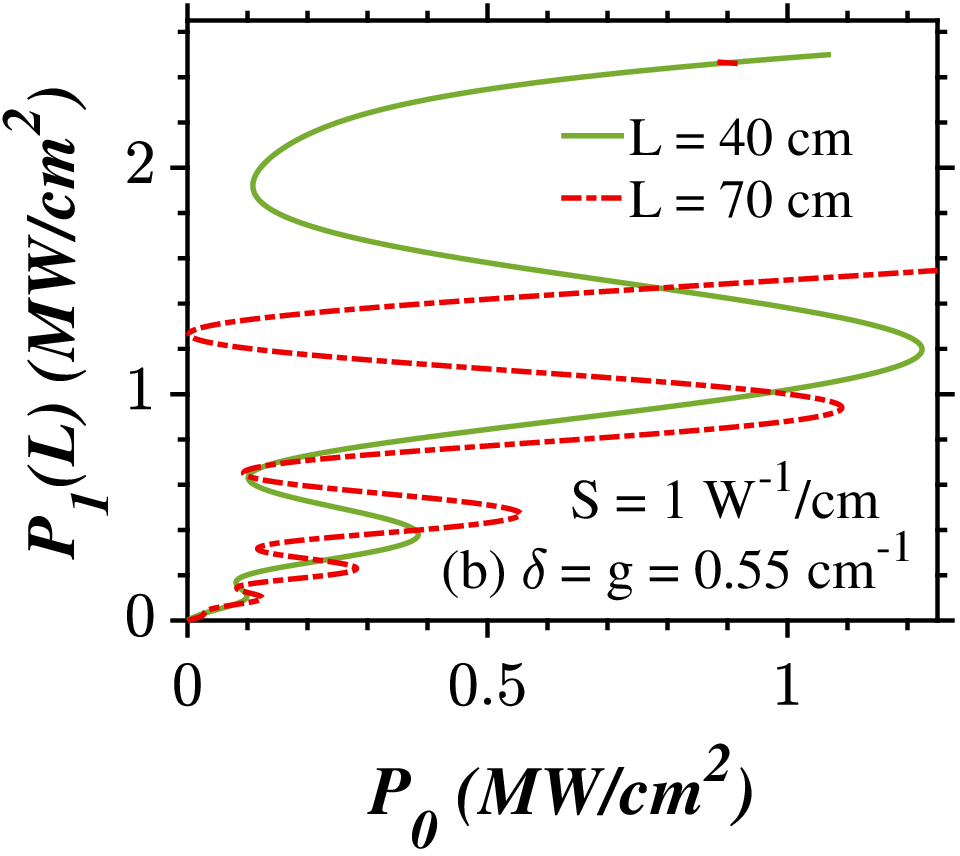}\\\includegraphics[width=0.5\linewidth]{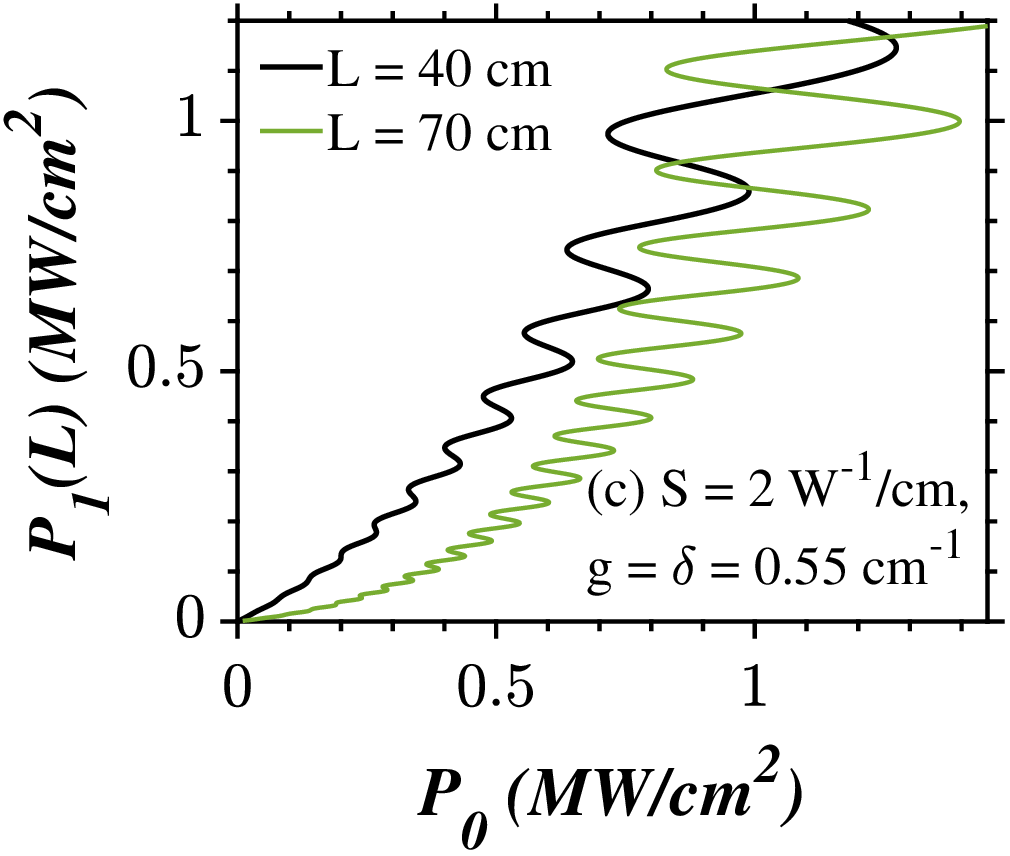}\includegraphics[width=0.5\linewidth]{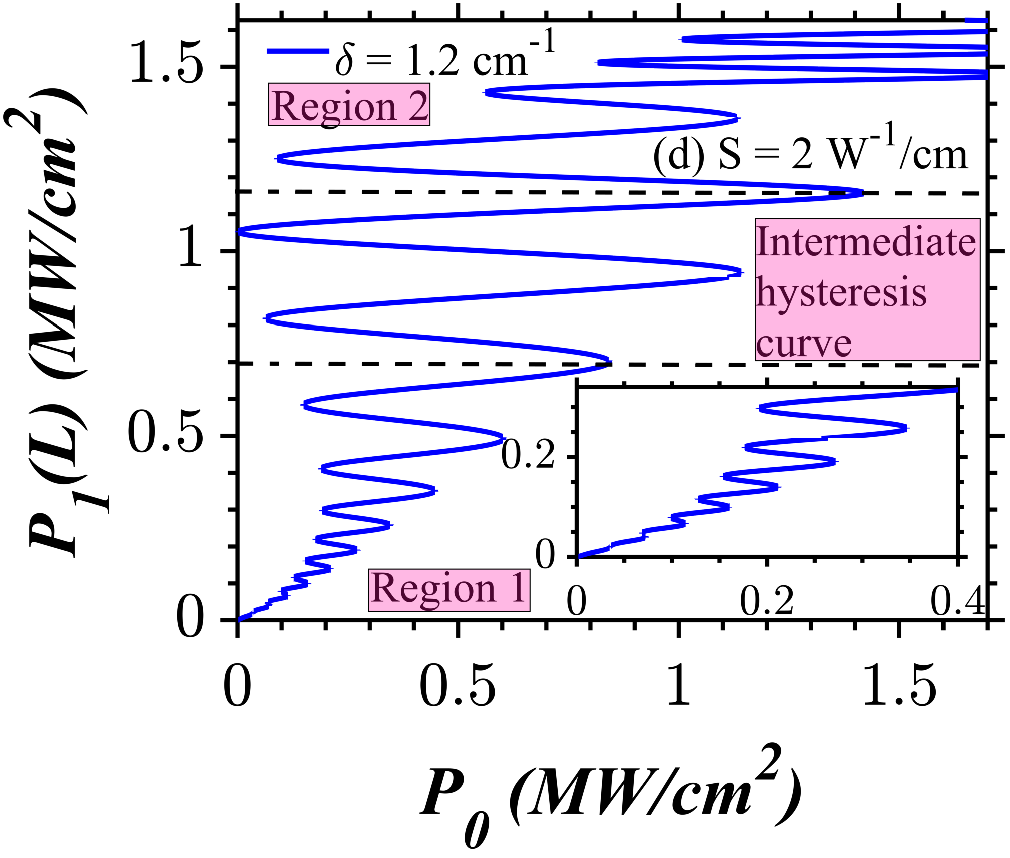}\\\includegraphics[width=0.5\linewidth]{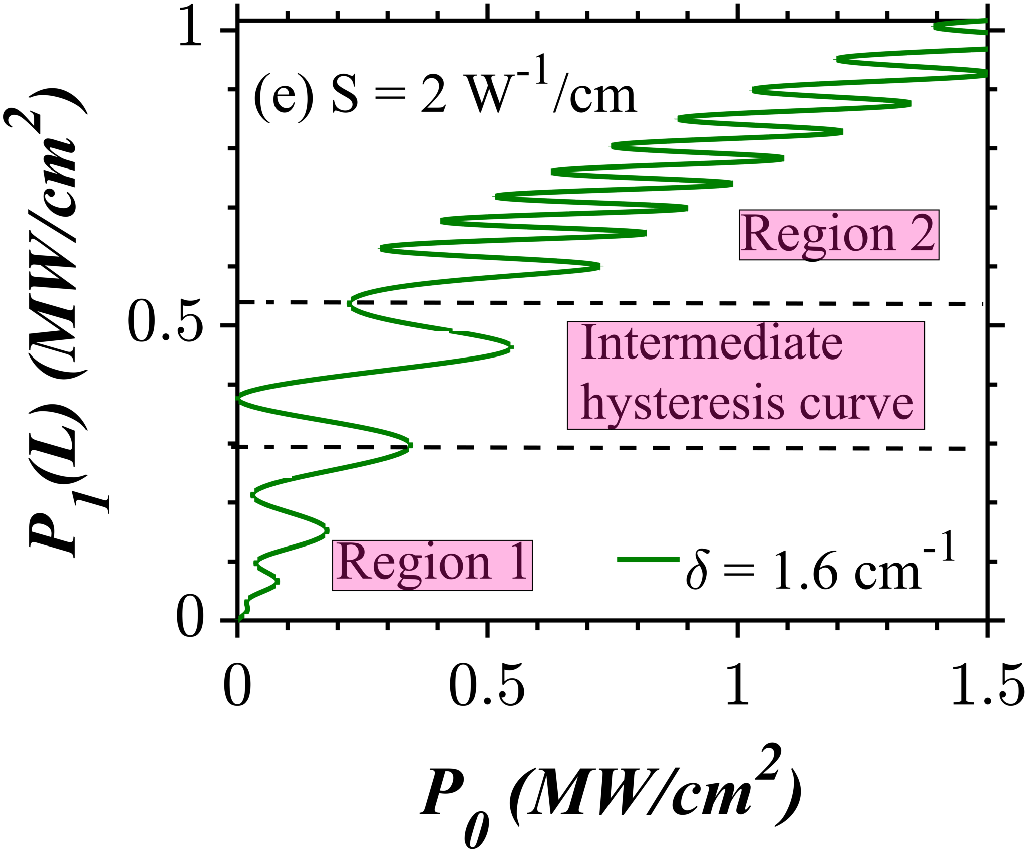}
	\caption{Increase in the number of hysteresis curves in a ramp-like OM curve exhibited by a broken PTFBG ($g = 0.55$ $cm^{-1}$) with an increase in the device length. The direction of light incidence is left. (a) -- (c) Ramp-like OM curve, (d) -- (e) mixed OM curve.}  
	\label{fig4}
\end{figure}
It is possible to alter the number of stable states by increasing the device length at fixed input intensity values, as shown in Fig. \ref{fig4}(a). 
 At $L = 20$, the feedback parameter is too small to support a number of desirable hysteresis curves. For instance, no switching occurs in the curves in Fig. \ref{fig1}(a) for $P_0 < 1.5$ $MW/cm^2$ at $\delta = 0$. When the device length increases, ramp-like OM curves emerge in Fig. \ref{fig4}(a) for the same value of input intensity. Multiple switching scenarios happen in a narrow range of input intensities under an increase in the device length to $L = 70$ $cm$.   In this fashion, the curves corresponding to $L = 70$ $cm$ feature a more number of hysteresis curves than $L = 20$ $cm$. As a result, the overall number of stable states admitted by the system also increases.  This kind of increase in the number of stable branches also happens under the frequency detuning and increasing the value of SNL as shown in Figs. \ref{fig4}(b) and (c). 

\subsubsection{Increase in number of stable states of mixed OM curves: $S = 2$ $W^{-1}/cm$}
\label{Sec:3E3}
We understand that the number of ramp-like and S-like hysteresis curves of the mixed OM curves increases with an increase in the device length, as shown in Figs. \ref{fig4}(d) and (e).
 When the value of input intensity is low, the ramp-like OM curves appear in the input-output characteristics and is marked as region 1.  In region 2, S-shaped OM curves appear on the top of the ramp-like hysteresis curve at higher intensities, as shown in Figs. \ref{fig4}(d) and (e). The width of the ramp-like hysteresis curves in region 1 increases with an increase in the input intensities. On the other hand, the hysteresis widths of the S-shaped hysteresis curves in region 2 are narrower than its former. The ramp-like OM curves appear predominantly  in Fig. \ref{fig4}(d) and the number of S-like hysteresis curves is significantly less than that of the ramp-like hysteresis curves. As we tune the value of the detuning parameter further, the number of ramp-like and S-like hysteresis curves decreases and increases, respectively, as shown in Figs. \ref{fig4}(e). 

An intermediate hysteresis curve separates the ramp-like (region 1) and S-shaped (region 2) regions of a mixed OM curve in Figs. \ref{fig4}(d) and (e). This hysteresis curve features a sharp increase and gradual variations in output intensities in its lower and upper branches, respectively. The other distinct feature that differentiates this intermediate hysteresis curve from the rest of the curves is that it features a switch-down mechanism at near-zero intensity. It is also inferred from Figs. \ref{fig4}(d) and (e) that under the frequency detuning mechanism, the switch-up intensity of the intermediate hysteresis curve (that shows a near-zero switch-down mechanism), ramp-like and S-like hysteresis curves get reduced.

\subsection{Role of gain and loss on the ramp-like and mixed OM curves}
\label{Sec:3E}
\begin{figure}
	\centering	\includegraphics[width=0.5\linewidth]{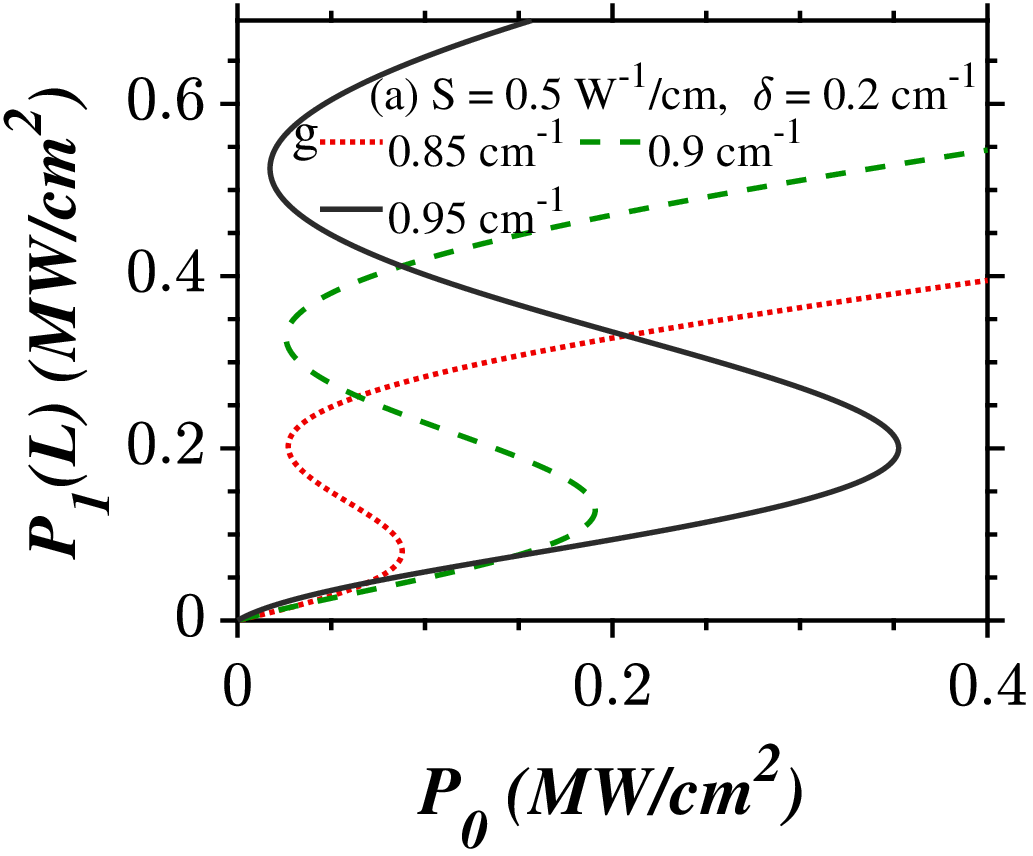}\includegraphics[width=0.5\linewidth]{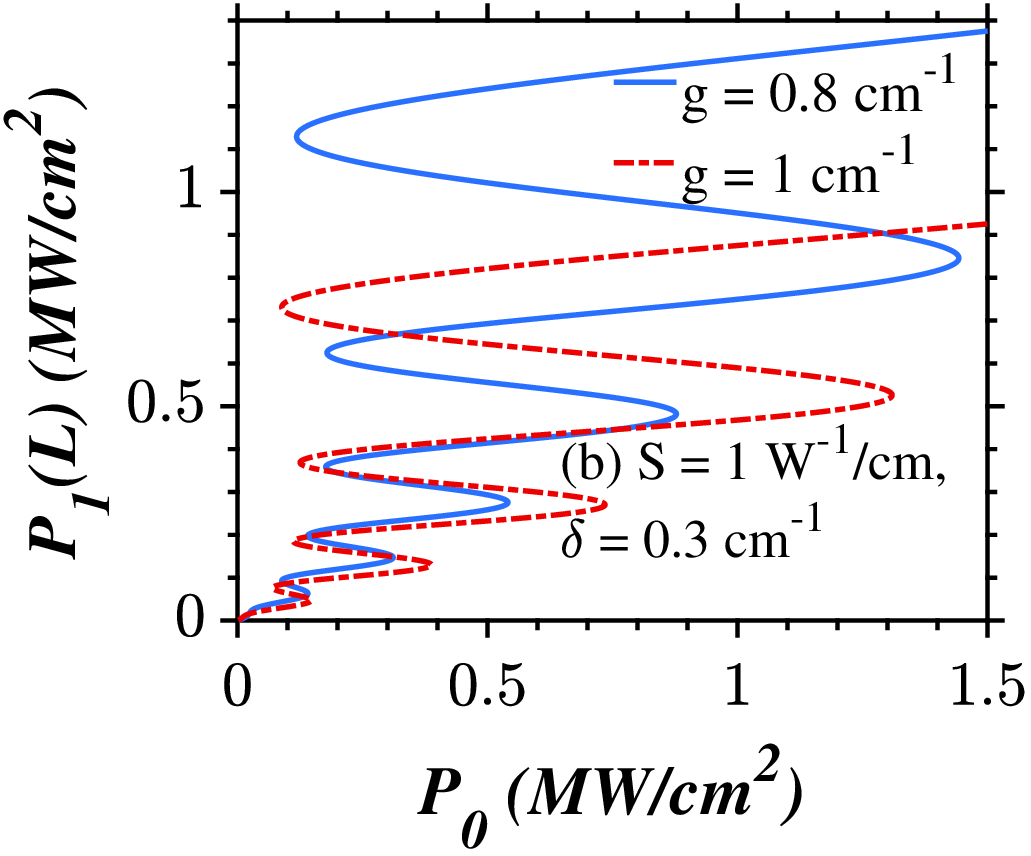}\\\includegraphics[width=0.5\linewidth]{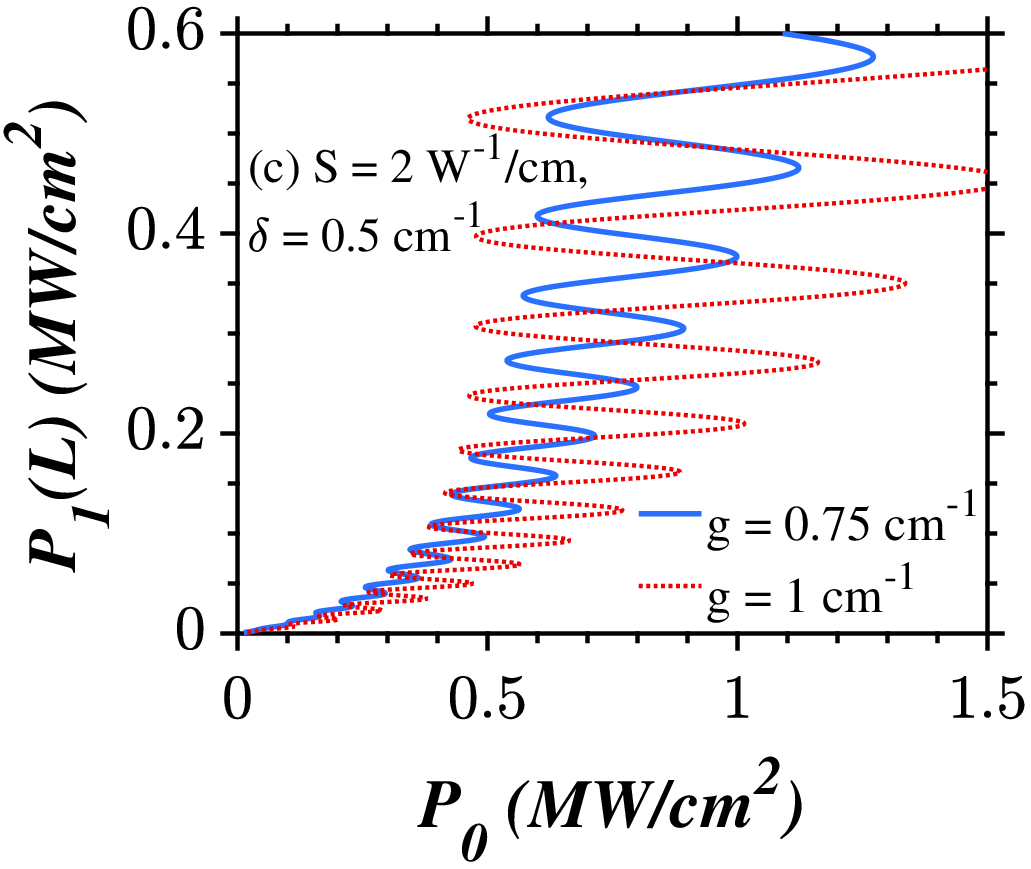}\includegraphics[width=0.5\linewidth]{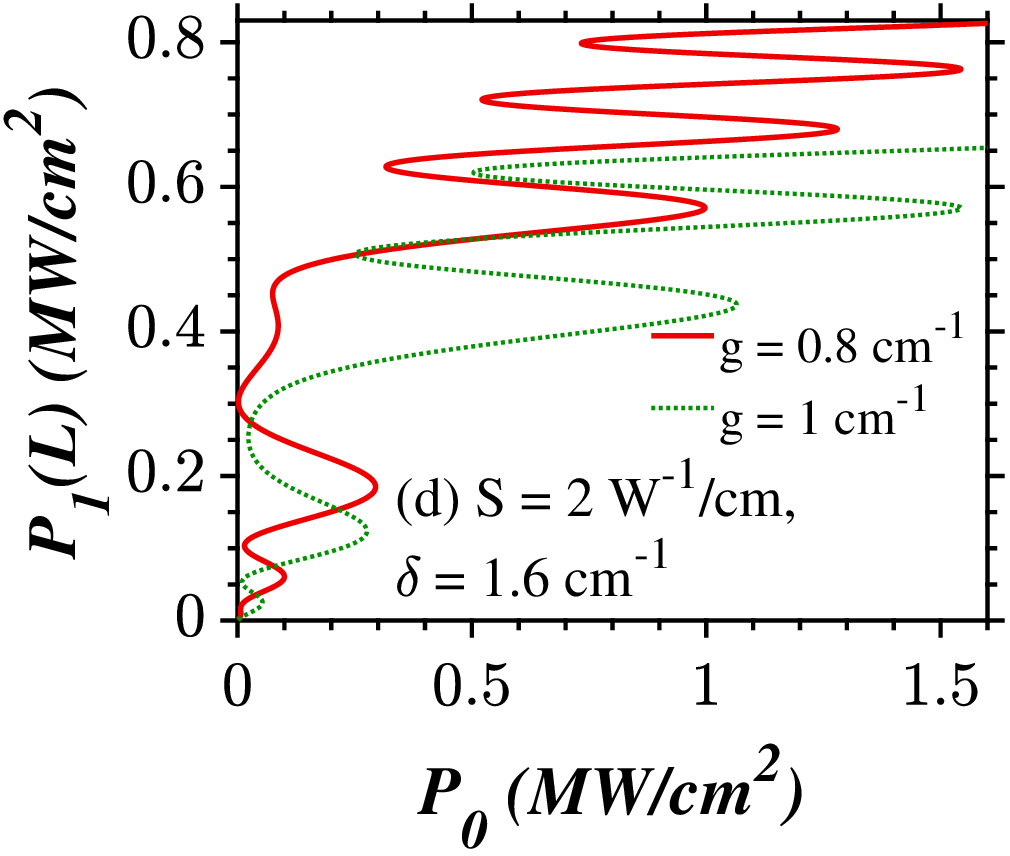}
	\caption{Increase in hysteresis width and switching intensities of different stable branches with an increase in gain and loss parameter at $L = 70$ cm. The light launching direction is left.}
	\label{fig5}
\end{figure}
  
An increase in the value of gain and loss parameter increases the switch-up and switch-down intensities of different stable branches in the OB (OM) curves, as illustrated in Fig. \ref{fig5}. This effect holds true for all the values of SNL and the detuning parameters, indicating that the gain and loss parameter increases the switching intensities for all the curves exhibited by the system, namely S-shaped, ramp-like, and mixed OM curves. Additionally, it results in an increase in the hysteresis width for all the stable branches of these curves. This contrasts with the results observed in the case of the present system operating in the unbroken $\mathcal{PT}$- symmetric regime, where it was found that increasing the gain and loss parameter is one of the viable options to reduce the switching intensities associated with OB (OM) curves admitted by the system. From these inferences, we can comment that the value of gain and loss parameter should be as close as possible to the value of the coupling coefficient and not too far away from it, irrespective of the operating regime, whether it is unbroken or broken $\mathcal{PT}$- symmetries. Any deviation from this condition leads to the undesirable effect increasing the switching intensities, which is what we need to consider critically while choosing the value of gain and loss parameter in the numerical simulations and practical experiments.

\section{OB in the broken $\mathcal{PT}$-symmetric regime: Right incidence}\label{Sec:4}
The gain and loss parameter did not play a role in reducing the switching intensities of the OM curves admitted by the PTFBG system with SNL in the broken regime, as illustrated in Fig. \ref{fig5}. Even though this effect is unfavorable, we cannot simply neglect the beneficial roles offered by the inclusion of gain and loss. For instance, the notion of $\mathcal{PT}$-symmetry opens a door for investigating the nonreciprocal switching in PTFBGs. Recall that the OB/OM curves generated by the system under left light incidence are noticeably distinct from those generated under right light incidence, particularly in terms of switching intensities. In the context of studying nonreciprocal switching in PTFBGs under the right light incidence, we also found that it serves as an alternate route to preserve the productive features of $\mathcal{PT}$-symmetry. Motivated by the fact that the amount of reduction in the switching intensities is appreciable by the condition of right light incidence in the PTFBG configurations than that of the left incidences which we have witnessed in our previous findings \cite{raja2019multifaceted,PhysRevA.100.053806,sudhakar2022inhomogeneous,sudhakar2022low,raja2022saturate}, we  would like to explore its usefulness for the same purpose in the presence of SNL and the broken $\mathcal{PT}$-symmetry.

\begin{figure}
	\centering	\includegraphics[width=0.5\linewidth]{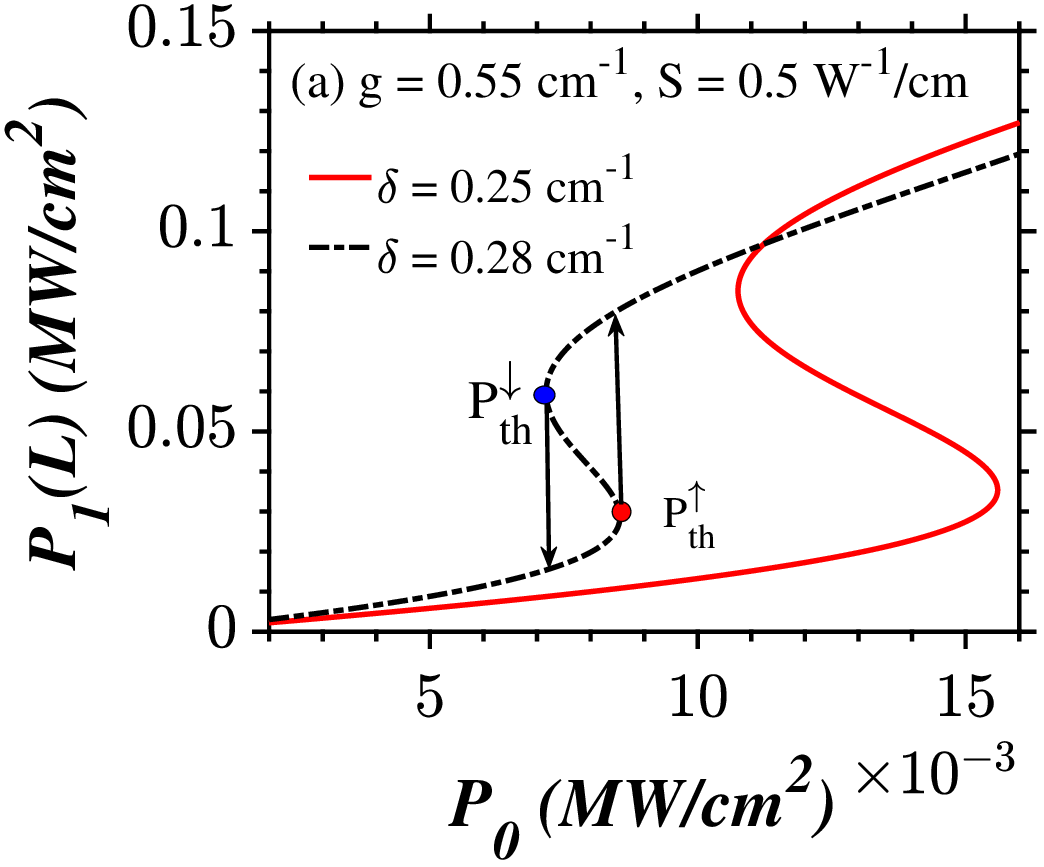}\includegraphics[width=0.5\linewidth]{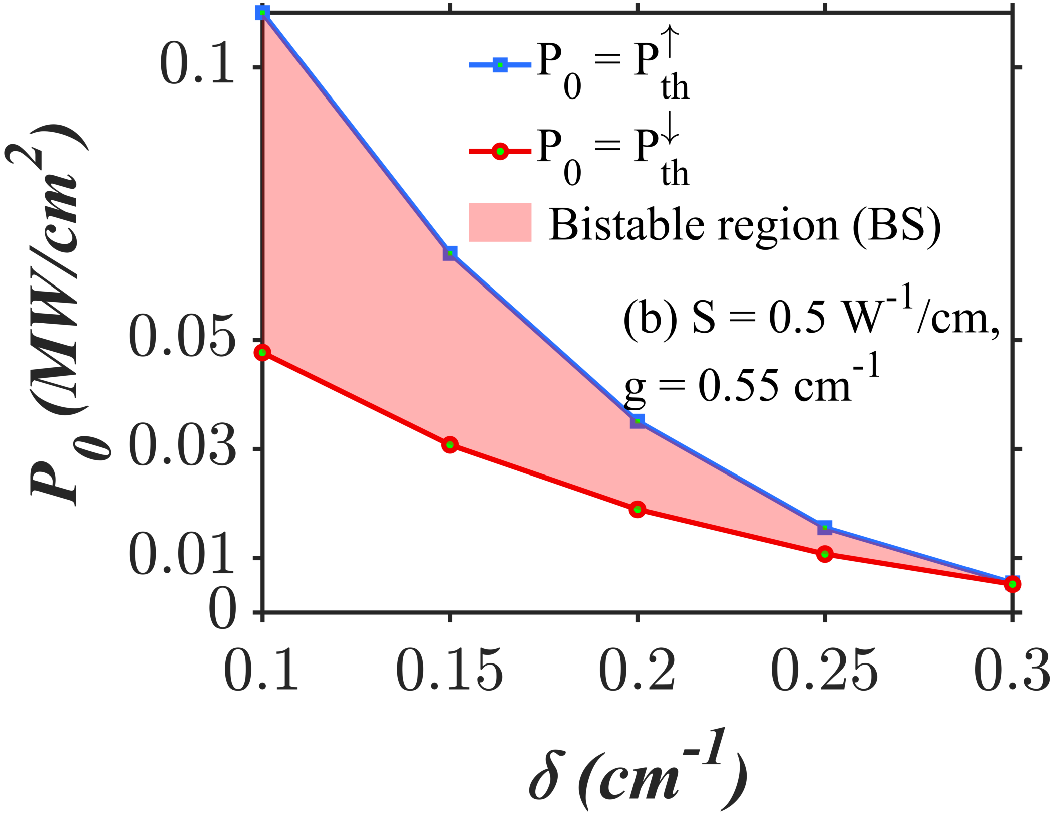}\\\includegraphics[width=0.5\linewidth]{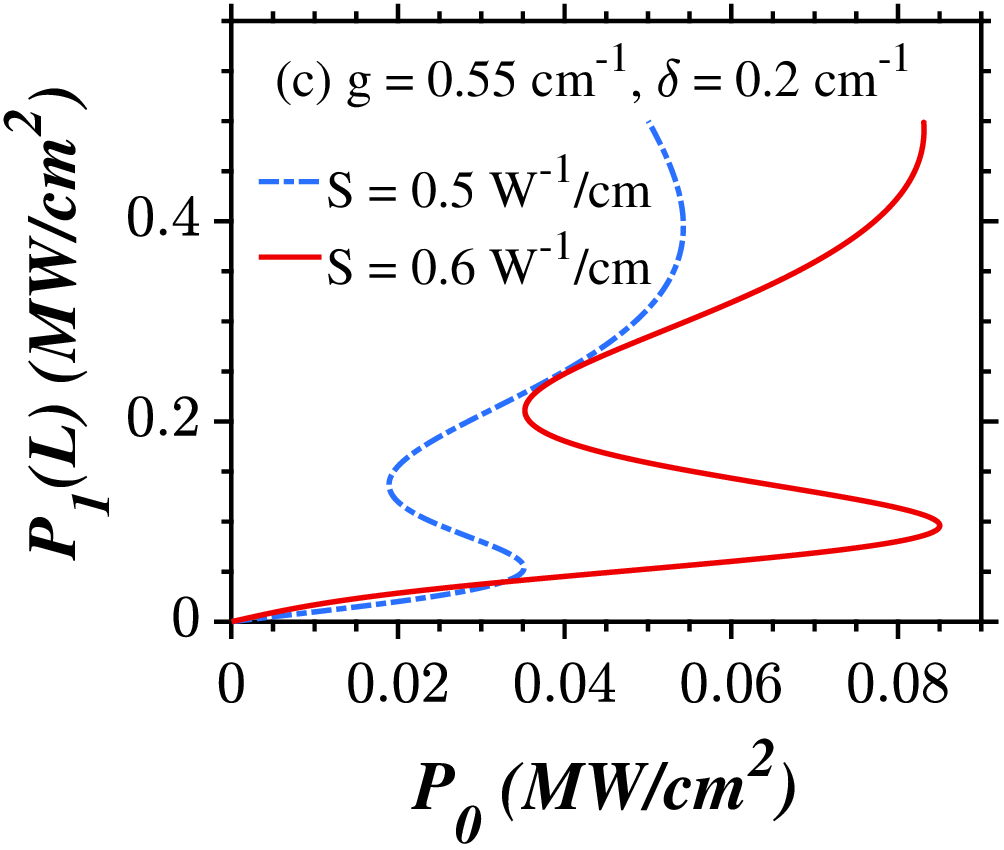}\includegraphics[width=0.5\linewidth]{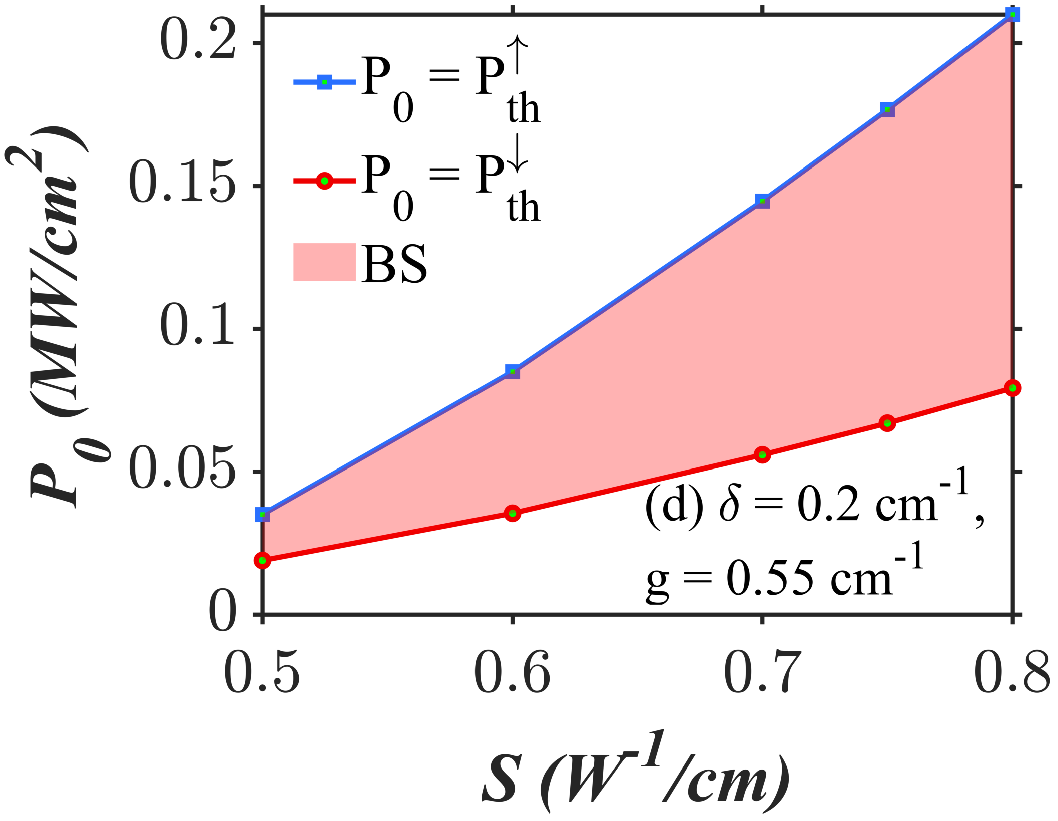}
	\caption{Role of (a)  detuning and (c) SNL parameter  on the S-shaped OB curves at $L = 20$ $cm^{-1}$. Variations in the switching intensities as a function of (b) detuning parameter at fixed values of $S$ and (d) SNL parameter at fixed values of $\delta$. The direction of light incidence is right. } 
	\label{fig6}
\end{figure}
\begin{figure}
	\centering	\includegraphics[width=0.5\linewidth]{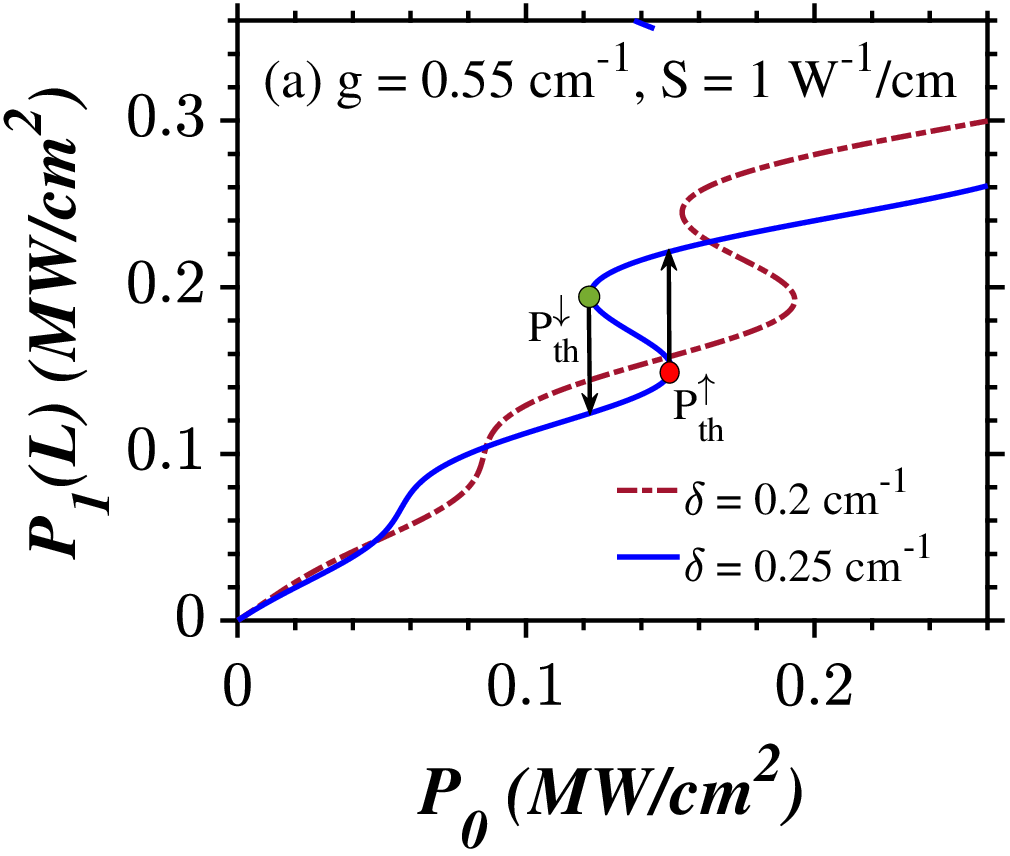}\includegraphics[width=0.5\linewidth]{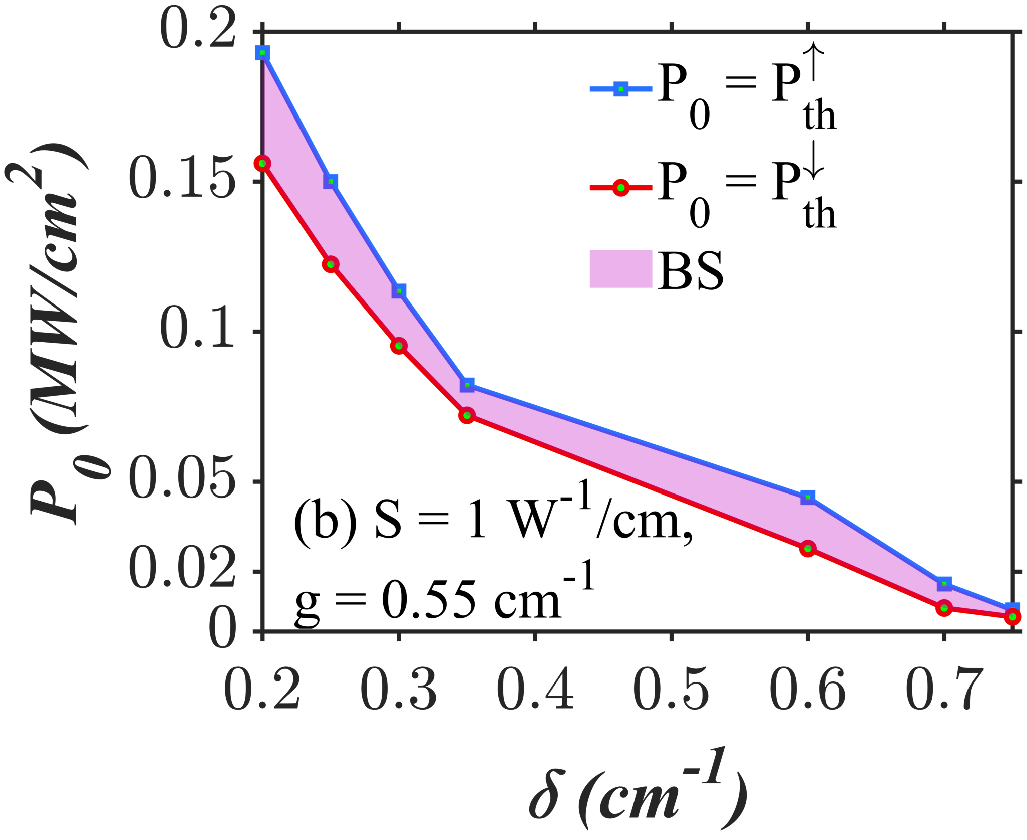}\\\includegraphics[width=0.5\linewidth]{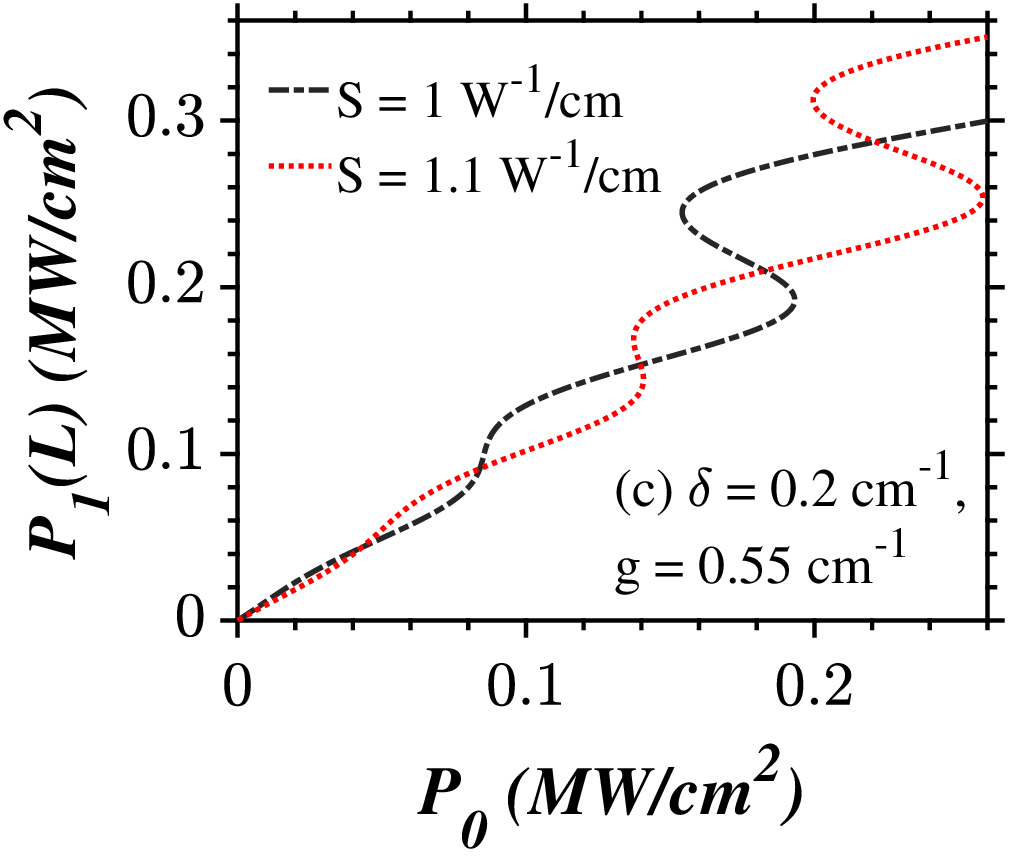}\includegraphics[width=0.5\linewidth]{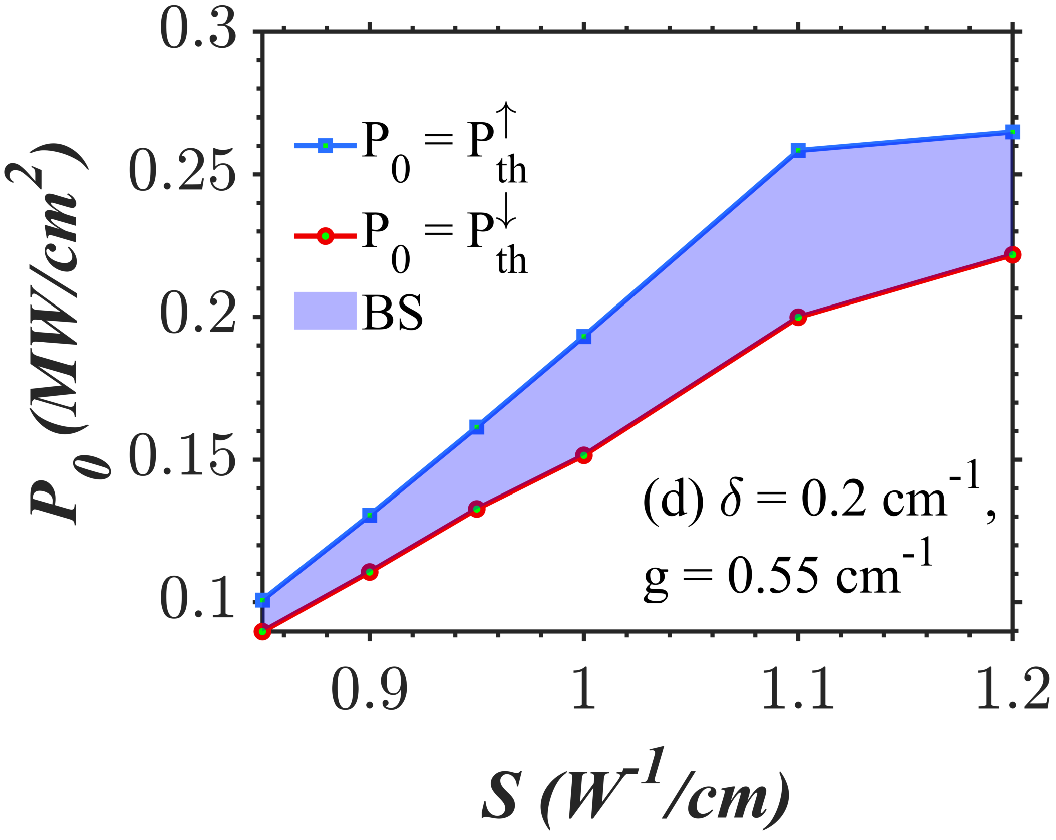}
	\caption{Role of (a)  detuning and (c) SNL parameter  on the ramp-like OB curves at $L = 20$ $cm^{-1}$. Variations in the switching intensities as a function of (b) detuning parameter at fixed values of $S$ and (d) SNL parameter at fixed values of $\delta$. The direction of light incidence is right. } 
	\label{fig7}
\end{figure}
\begin{figure}
	\centering	\includegraphics[width=0.5\linewidth]{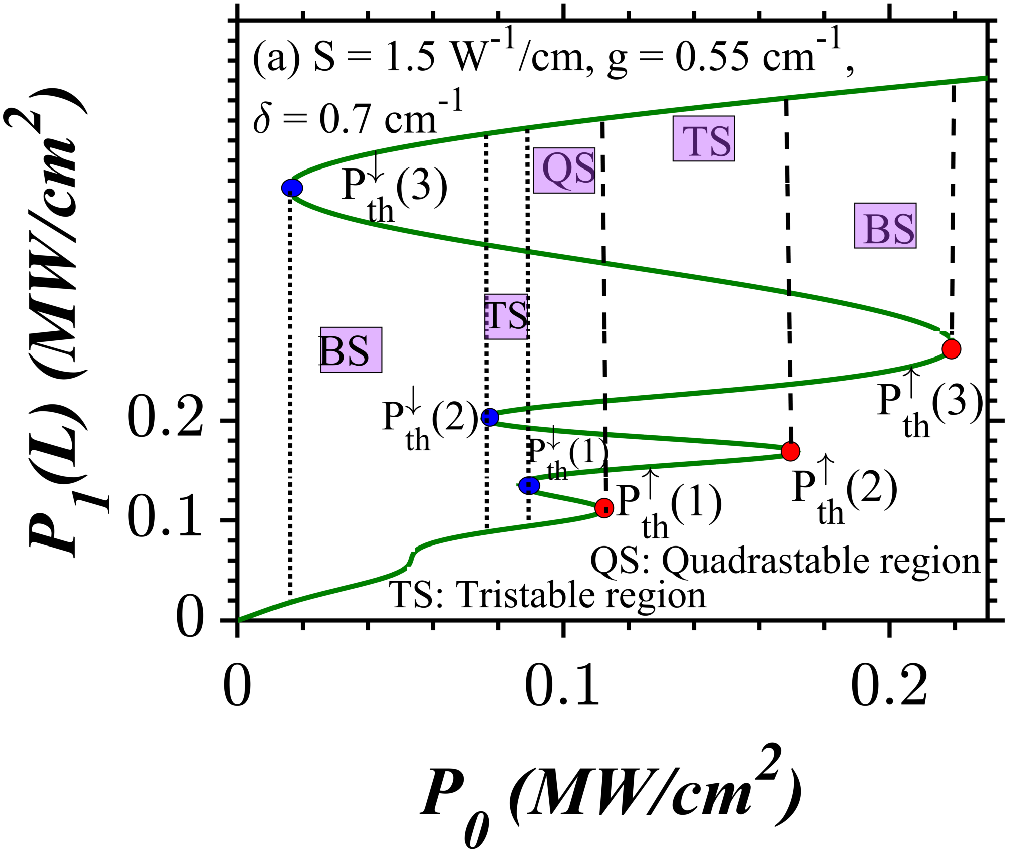}\includegraphics[width=0.5\linewidth]{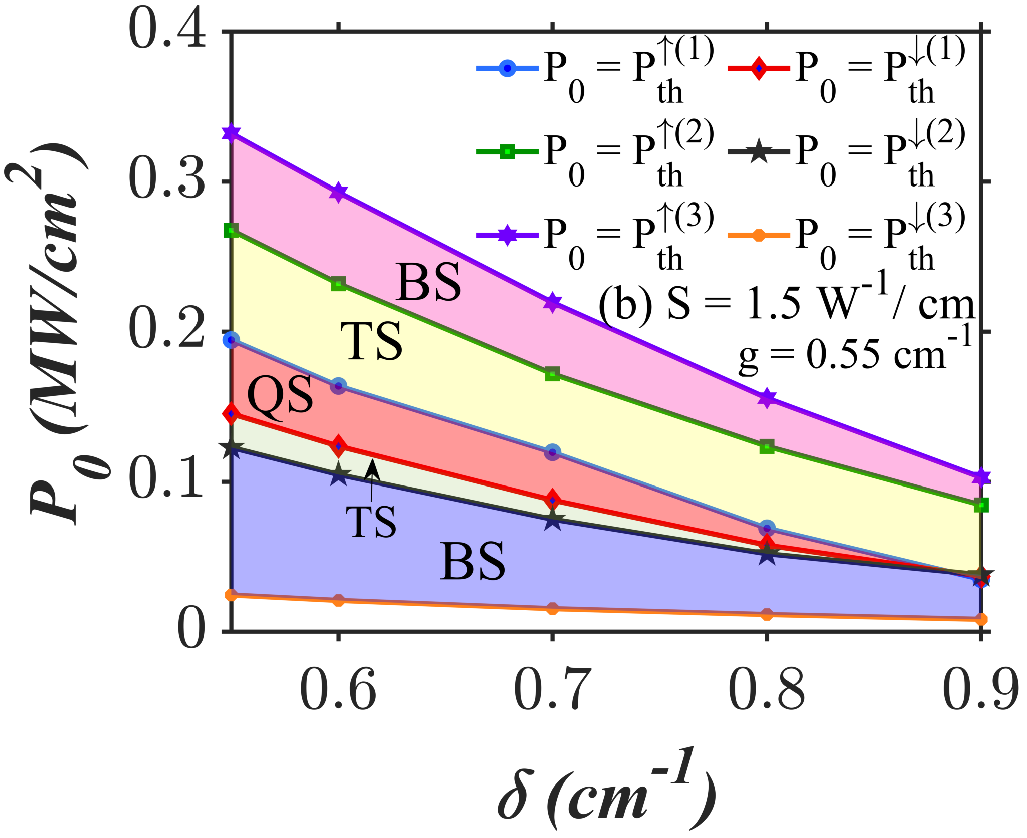}\\\includegraphics[width=0.5\linewidth]{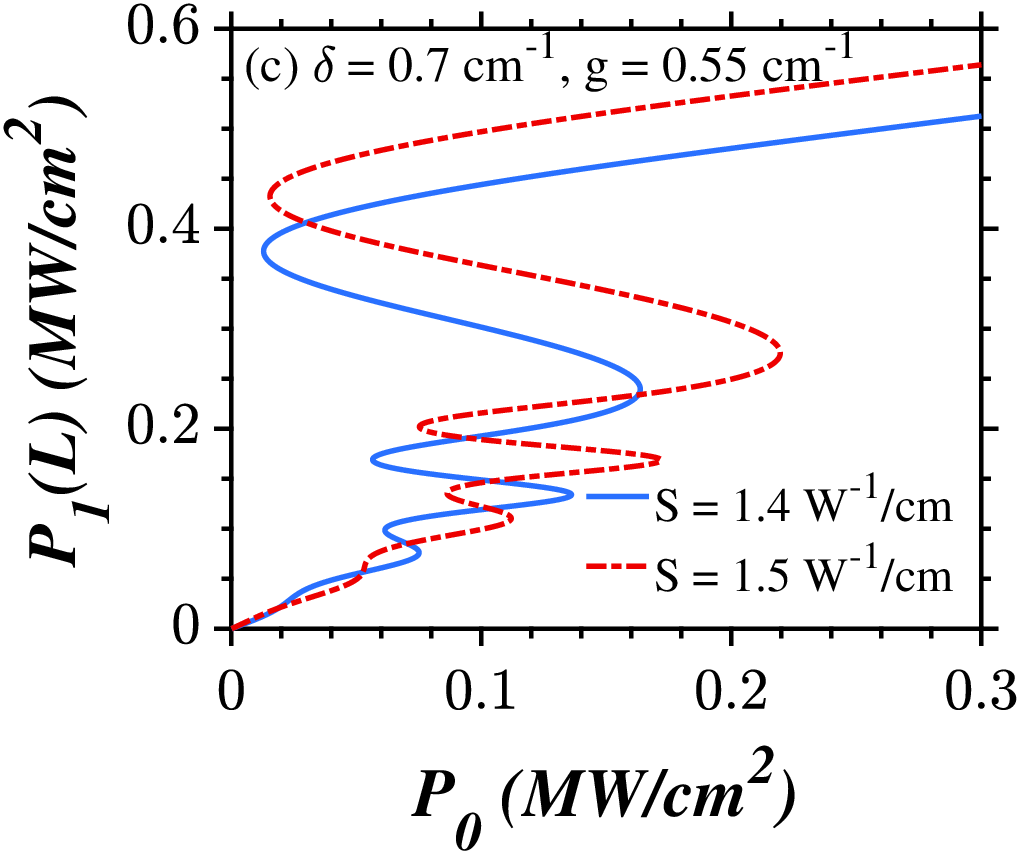}\includegraphics[width=0.5\linewidth]{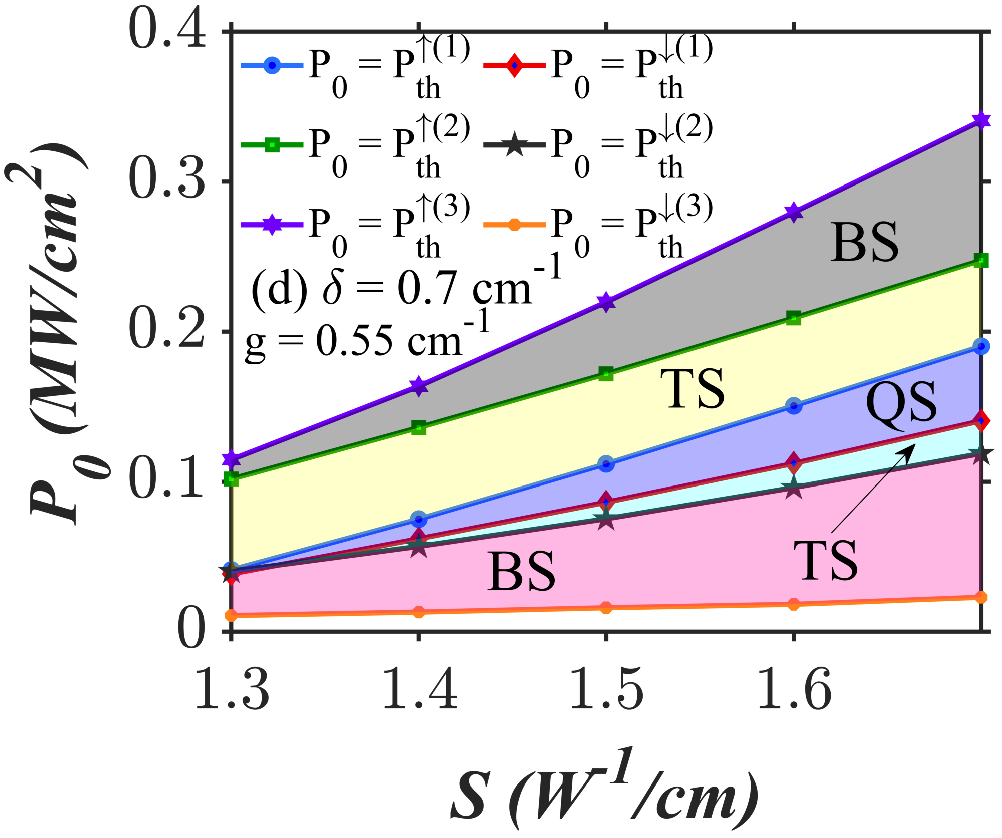}
	\caption{ Role of (a)  detuning and (c) SNL parameter  on the ramp-like OM curves at $L = 20$ $cm^{-1}$. Also, (a) depicts various stable regions namely bistable (BS), tristable (TS), quadrastable (QS) featuring in a ramp-like OM. Variations in the switching intensities as a function of (b) detuning parameter at fixed values of $S$ and (d) SNL parameter at fixed values of $\delta$.  The direction of light incidence is right.  } 
	\label{fig8}
\end{figure}
\begin{figure}
	\centering	\includegraphics[width=0.5\linewidth]{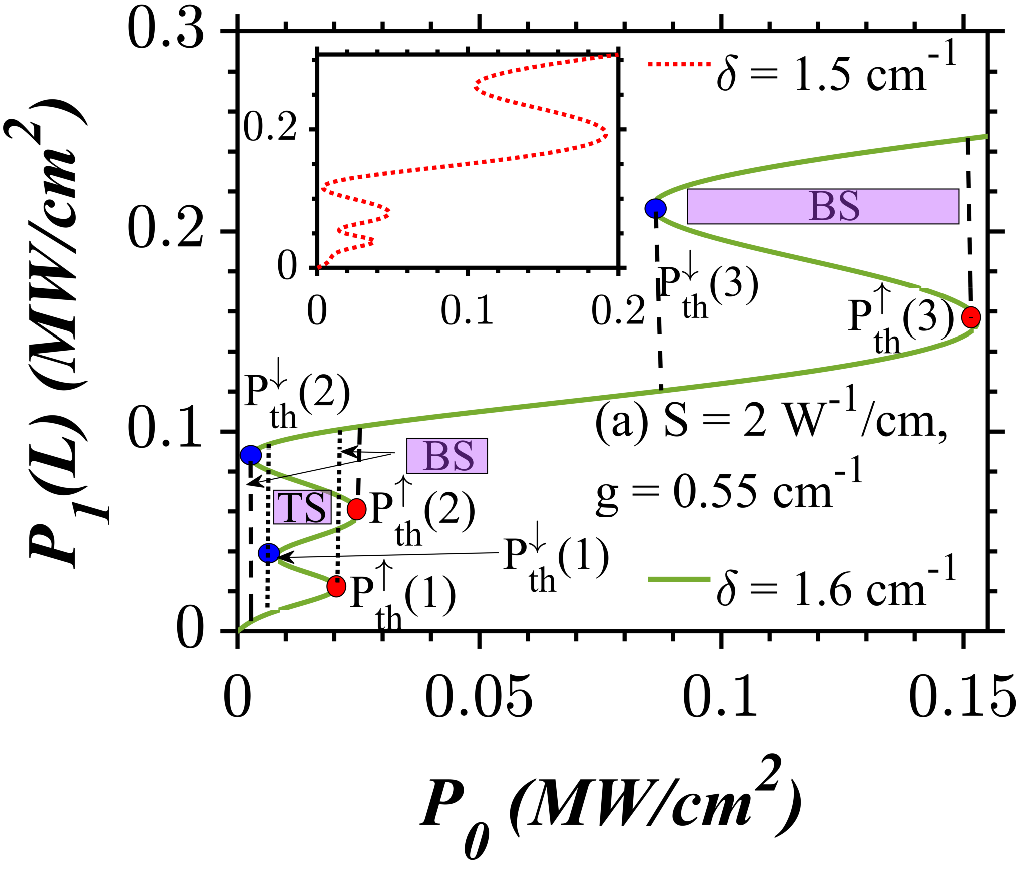}\includegraphics[width=0.5\linewidth]{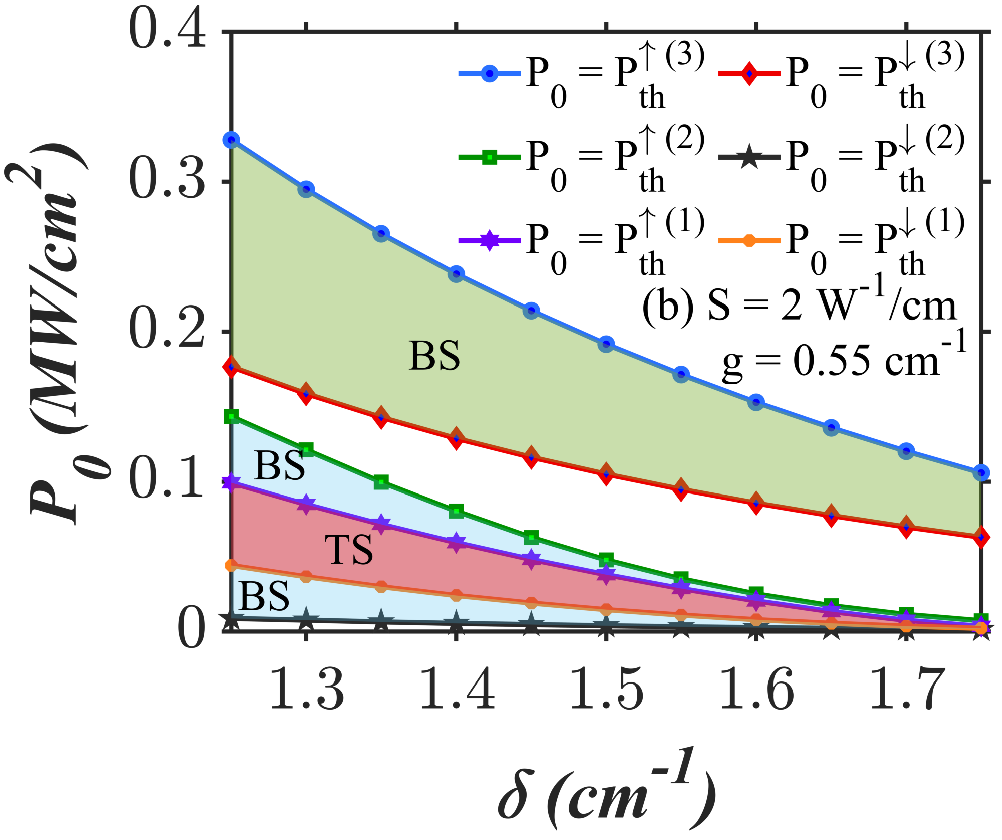}\\\includegraphics[width=0.5\linewidth]{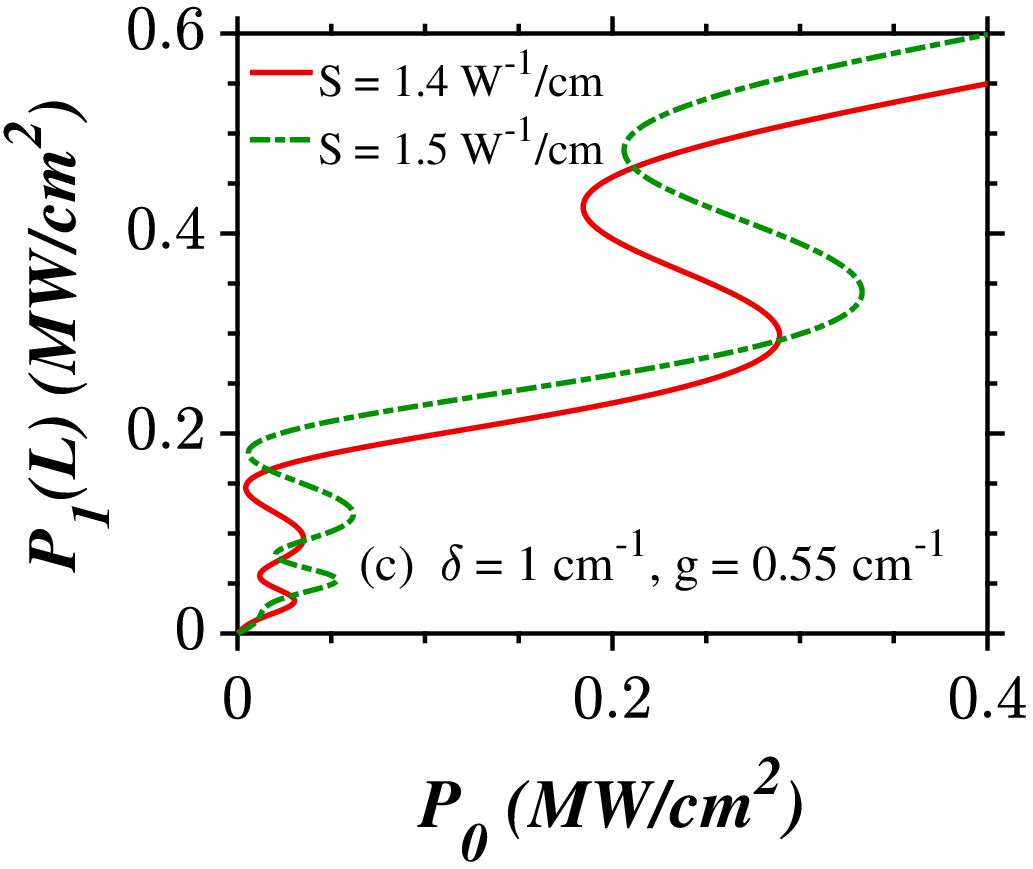}\includegraphics[width=0.5\linewidth]{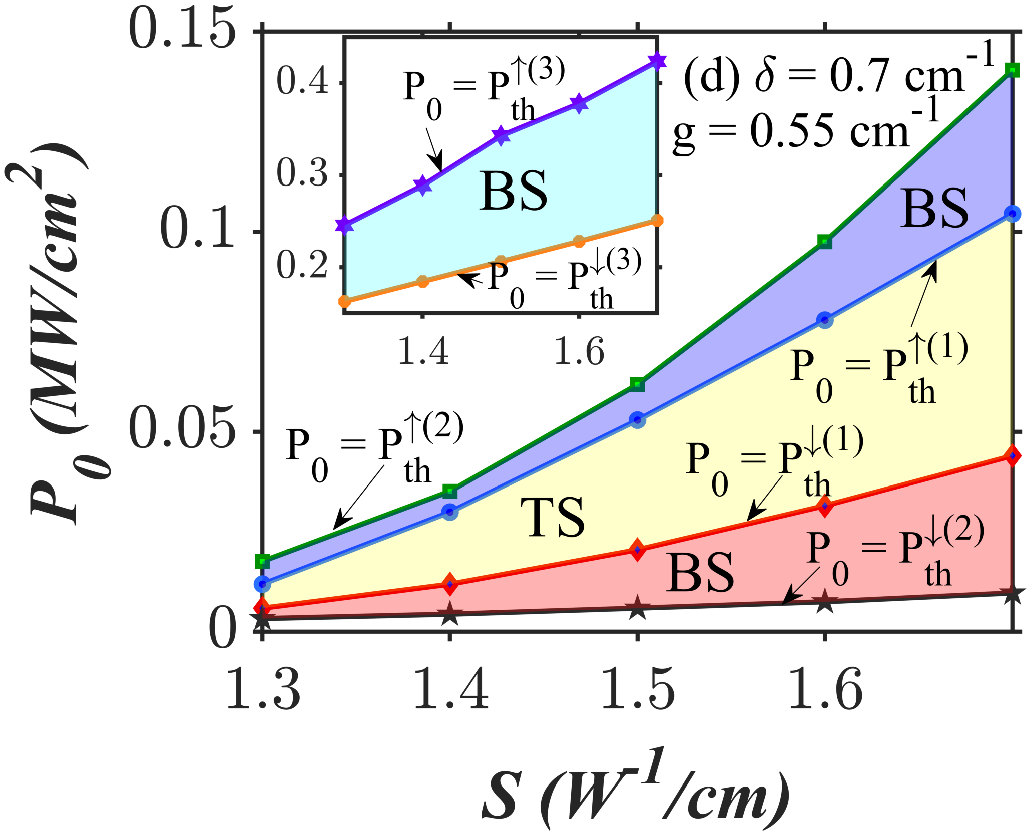}
	\caption{ Role of (a)  detuning and (c) SNL parameter  on the mixed OM curves at $L = 20$ $cm^{-1}$. Also, (a) depicts various stable regions featuring in a mixed OM. Variations in the switching intensities as a function of (b) detuning parameter at fixed values of $S$ and (d) SNL parameter at fixed values of $\delta$.  The direction of light incidence is right.  } 
	\label{fig9}
\end{figure}

\subsection{Low power S-shaped OB curves  at $L = 20$ cm}\label{Sec:4A}
Figure \ref{fig6}  reveals that reversing the direction of light incidence notably decreases the switching intensities of the S-shaped OB curves generated by the broken PTFBG with SNL. The observed result is a direct outcome of the maxima of the total optical field being positioned within the gain regions under the right light incidence condition \cite{kulishov2005nonreciprocal}, a characteristic shared among all the nonlinear PTFBG configurations examined to date. The switching intensities and the hysteresis width decrease with an increase in the detuning parameter, as shown in Fig. \ref{fig6}(a). This fact is once again confirmed through a map of variation of switching intensities pertaining to the S-shaped OB curves as a function of continuously varying detuning parameter at fixed values of SNL term, as shown in Fig. \ref{fig6}(b). On the other hand, if the values of the SNL parameter are increased at fixed values of the detuning parameter, it produces an undesirable effect in the form of an increase in the switching intensities, as shown in Fig. \ref{fig6}(c), which is also confirmed in Fig. \ref{fig6}(d), drawn in a parametric plot of critical power versus the saturable nonlinearity for its wide range of values. 
\subsection{Low power ramp-like OB curves  at $L = 20$ cm}\label{Sec:4B}
As the value of saturable nonlinearity is increased, a transformation from the S-shaped OB to the ramp-like OB is inevitable, irrespective of the change in the light launching direction from left to right, as confirmed by the plots depicted in Fig. \ref{fig7}. An increase in the detuning parameter results in a reduction of both switching intensities and hysteresis width, as shown in Figs. \ref{fig7}(a) and (b). In contrast, increasing the value of SNL parameter with the detuning parameter held at constant leads to a rise in the switching intensities, as delineated in Figs. \ref{fig7}(c) and (d).

\subsection{Low power ramp-like OM curves at $L = 20$ cm}\label{Sec:4C}
When the value of nonlinearity is further increased (at fixed values of input intensities) in the simulations, the input-output characteristics curve will begin to support a greater number of stable states, as depicted in Fig. \ref{fig8}. The width of the successive hysteresis curves, emerging above the initial ramp-like stable branch, expands with tuning the input intensity. The output intensity varies sharply in all branches, leading to the formation of ramp-like OM curves. With a further increase in the frequency detuning, the switching intensities needed to transition from the first ramp-like stable state to the second decrease significantly, as illustrated in Figs. \ref{fig8}(a) and (b). On a contrary, in this case too, increasing the SNL parameter has a undesirable impact on the ramp-like OM curve, leading to an increase in the switching intensities, as illustrated in Figs. \ref{fig8}(c) and (d). 

\begin{table}[htbp]
	\caption{Number of stable states vs $P_0$ in Figs. \ref{fig8}(b),(d), \ref{fig9}(b),(d). }
	
	\begin{center}
		\begin{tabular}{c c c}
			\hline \hline
			{Number of}& {Ramp-like OM}& {Mixed OM} \\
			{stable states}&{Figs. \ref{fig8}(b) and (d) }&{Figs. \ref{fig9}(b) and (d)} \\
	 	\hline

			{2}&{$P_{th}^{\downarrow(3)}$ $\le$ $P_0$ $<$ $P_{th}^{\downarrow(2)}$}&{ $P_{th}^{\downarrow(2)}$ $\le$ $P_0$ $<$ $P_{th}^{\downarrow(1)}$} \\
			
			{}&$P_{th}^{\uparrow(2)}$ $\le$ $P_0$ $<$ $P_{th}^{\uparrow(3)}$& $P_{th}^{\uparrow(1)}$ $<$ $P_0$ $\le$ $P_{th}^{\uparrow(2)}$ \\
			
			{}&{--}& $P_{th}^{\downarrow(3)}$ $<$ $P_0$ $\le$ $P_{th}^{\uparrow(3)}$ \\
			{}&{}&{}\\

			{3}&$P_{th}^{\downarrow(2)}$ $\le$ $P_0$ $<$ $P_{th}^{\downarrow(1)}$& {$P_{th}^{\downarrow(1)}$ $\le$ $P_0$ $<$ $P_{th}^{\uparrow(1)}$} \\
			{}&$P_{th}^{\uparrow(1)}$ $\le$ $P_0$ $<$ $P_{th}^{\uparrow(2)}$& {--} \\
				{}&{}&{}\\

			{4}&{$P_{th}^{\downarrow(1)}$ $\le$ $P_0$ $<$ $P_{th}^{\uparrow(1)}$}&{--} \\
			\hline\hline
		\end{tabular}
	
		\label{tab7}
	\end{center}

\end{table}

As mentioned in Table \ref{tab7}, the output of the system is tristable (TS) and quadrastable (QS), which means that there exist three and four output states, respectively for a given value input of input intensity, as shown in Figs. \ref{fig8}(b) and (d). Also, these two plots and Table \ref{tab7} illustrate that the system may (also) remain bistable for a certain range of input intensities.

\subsection{Low power mixed OM curves at $L = 20$ cm}\label{Sec:4D}	
	For larger values of the detuning parameter, the system gives rise to the formation of mixed OM curves, as shown in Fig. \ref{fig9}. At low intensities, the OB curve is characterized by a ramp-like shape, transitioning to an S-shaped curve at higher intensities. Importantly, mixed OM curves in Fig. \ref{fig9} under right light incidence occur at lower intensities than the mixed OM curve shown in Fig. \ref{fig3}(b) under the left light incidence.
	
Similar to other types of OB and OM curves, the switching intensities associated with mixed OM curves decrease with an increase in the detuning parameter and increase with an increase in the SNL parameter, as illustrated in Figs. \ref{fig9}(a) and (c), respectively. The system is bistable everywhere, except when the input intensity falls in the range $P_{th}^{\downarrow(1)}\le P_0 \le P_{th}^{\uparrow(1)}$, where the system supports tristability, as confirmed by Figs. \ref{fig9}(b) and (d). It is important to note that the ranges of SNL and detuning parameters, leading to the emergence of S-shaped OB, ramp-like OB, ramp-like OM, and mixed OM in Figs. \ref{fig6} -- \ref{fig10}, are broad and not discrete ones. 
	
		\begin{table}[h!]
		\caption{Role of different control parameters on the critical switch up ($P_{th}^{\uparrow}$), down ($P_{th}^{\uparrow}$) intensities, hysteresis width
				($\Delta^{\uparrow \downarrow}$), and number of stable states pertaining to the hysteresis curves shown by PTFBG with SNL in broken $\mathcal{PT}$- symmetric regime, as inferred from Figs. 1 -- 9.} 
		\begin{center}
	
				\begin{tabular}{c c c c c}
					\hline
					\hline
					{Fig. no.}&	{Increase in}&{Impact on} & {Impact on}&{Impact on }  \\
					{}&	{the control}&{$P_{th}^{\uparrow}$ and} & {$\Delta^{\uparrow\downarrow}$}&{the number}  \\
					{}&		{parameter}&{$P_{th}^{\downarrow}$} & {}&{of stable}  \\
					{}&			{}&{}& {}&{states} \\
					
					\hline
					{Figs. }&		{$\delta$}&{decreases}& {decreases}&{insignificant}  \\
					{1 -- 3 and 7-- 9}&		{}&{}& {}&{changes}  \\
					{}&{}&{}&{}&{}\\
					{Figs. }&			{$S$}&{increases}& {increases}&{moderately}  \\
					{1 -- 3 and 7 -- 9}&				{}&{}& {}&{increases}  \\
					{}&{}&{}&{}&{}\\
					{Fig. 4}&				{$L$}&{decreases}&{increases}&{increases} \\
					{}&{}&{}&{}&{}\\
					{Fig. 5}&				{$g$}&{increases}&{increases}&{moderately} \\
					{}&			{}&{}& {}&{decreases} \\

					\hline\hline
					
				\end{tabular}

			\label{tab9}
		\end{center}
	\end{table}

\subsection{Low power ramp-like OM curves with vortex-like envelope at $L = 70$ cm and $\delta < 1$     $cm^{-1}$}
\label{Sec:4E}
Previously, we found that the curves simulated at $S = 2$ $W^{-1}/cm$ and $L = 70$ $cm$ feature more stable states in a given range of input intensities in Figs. \ref{fig4}(c) -- (e). What makes Figs. \ref{fig4}(d) and (e) even more fascinating is the transition of the nature of the hysteresis curves from ramp-like to mixed OM curves with near-zero switch-down intensities under the frequency detuning. The input intensities required to realize these states are slightly higher in these figures.   We know that launching the light from the rear end of the PTFBG aids in reducing the switching intensities \cite{raja2019multifaceted,PhysRevA.100.053806,sudhakar2022inhomogeneous,sudhakar2022low}. This motivated us to investigate these  OM curves in Fig. \ref{fig4}(c) -- (e) under a reversal in the direction of light incidence.

\begin{figure}
	\centering	\includegraphics[width=0.5\linewidth]{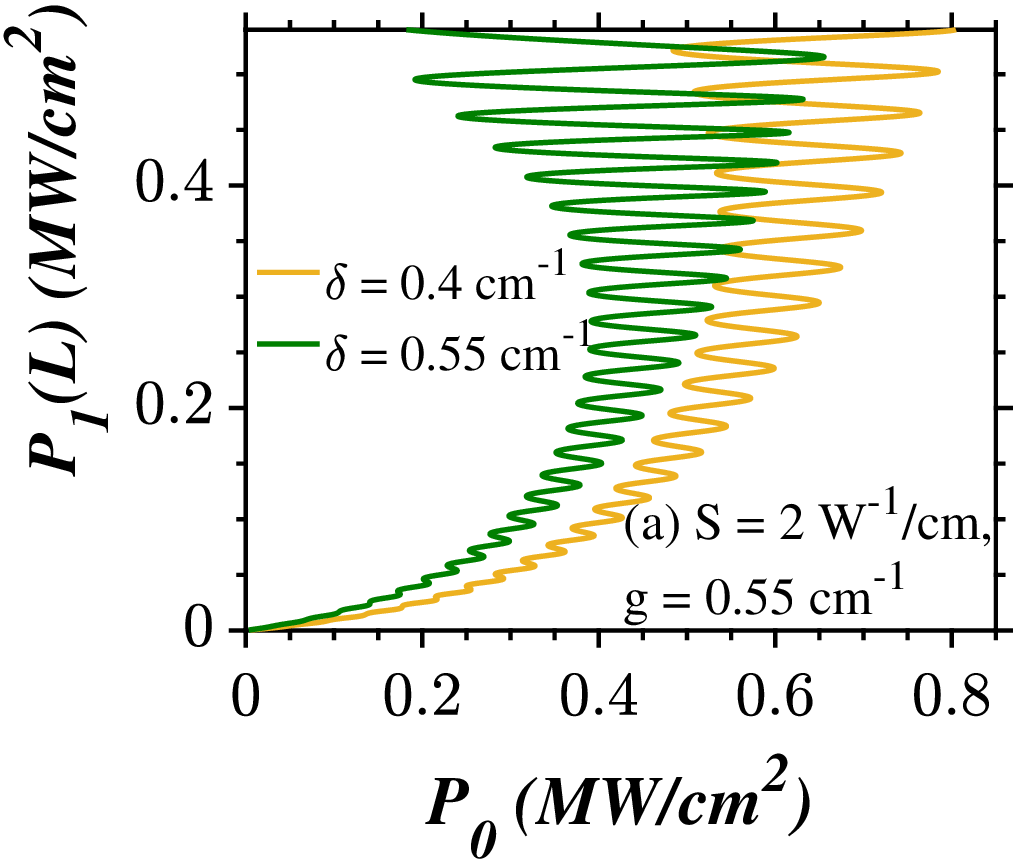}\includegraphics[width=0.5\linewidth]{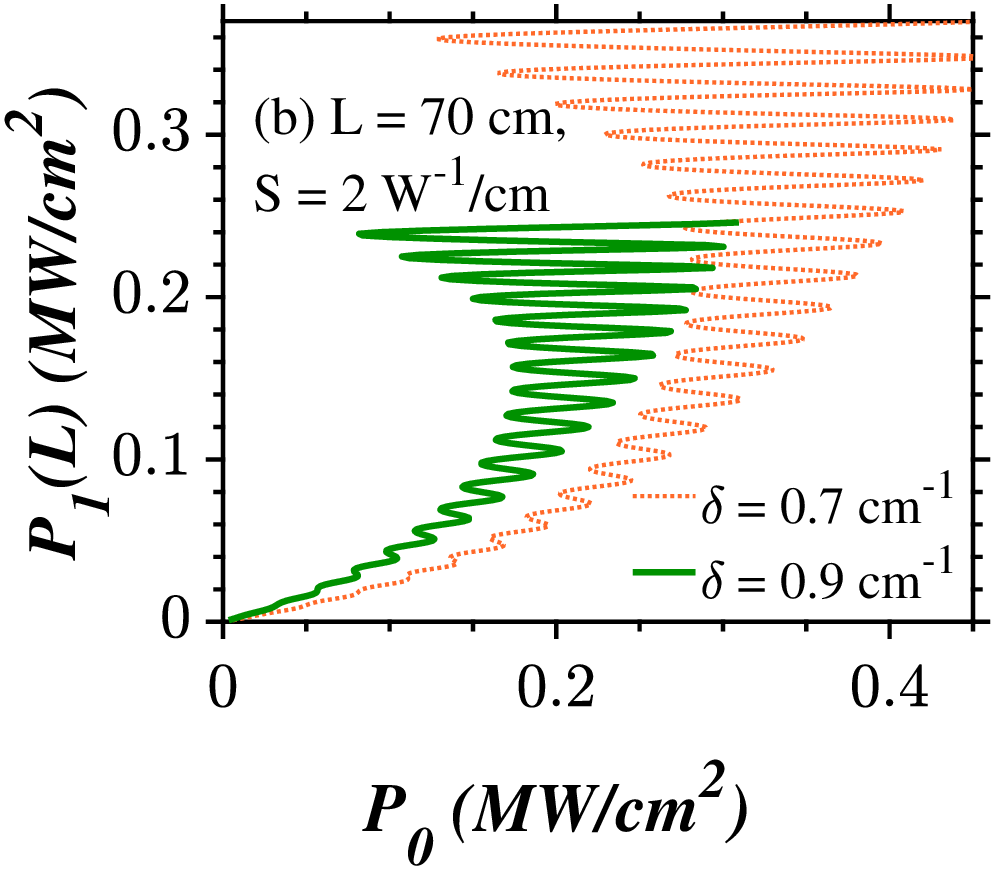}
	\caption{Formation of low-power ramp-like OM with vortex-like envelope in the input-output characteristics of a broken PTFBG with SNL at $L = 70$ $cm$ and $S = 2$ $W^{-1}/cm$. The direction of light incidence is right. }
	\label{fig10}
\end{figure}

\begin{figure}
	\centering	\includegraphics[width=0.5\linewidth]{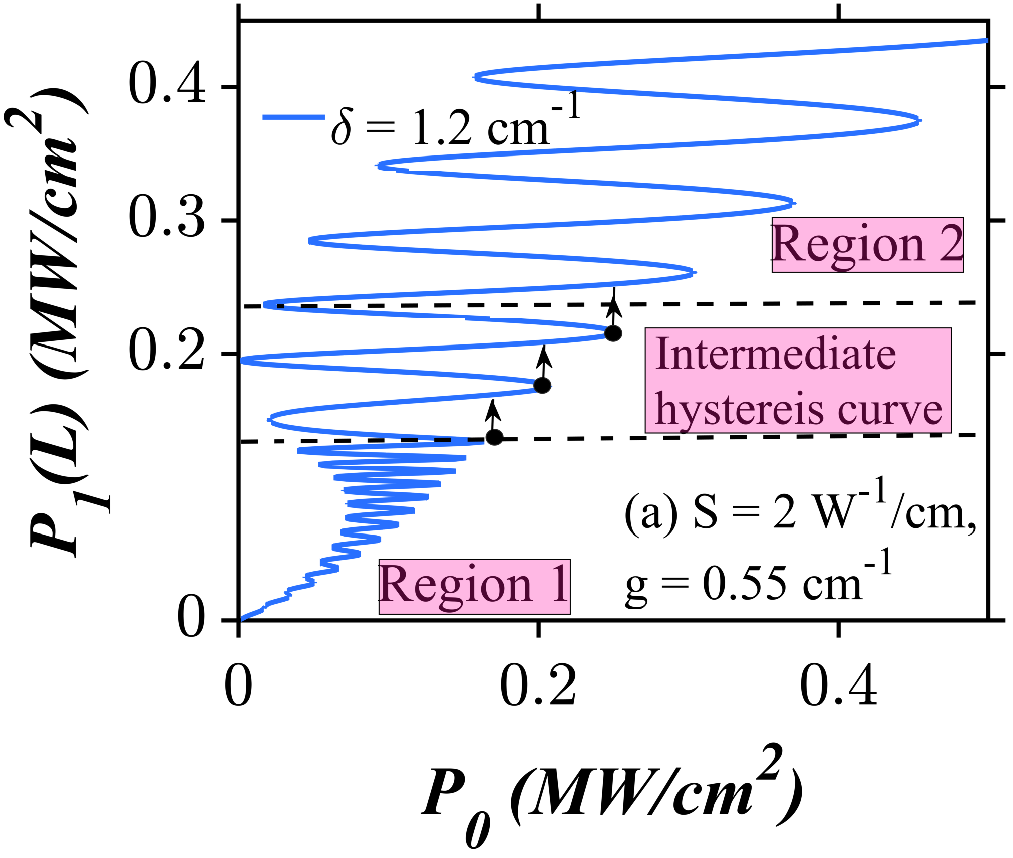}\includegraphics[width=0.5\linewidth]{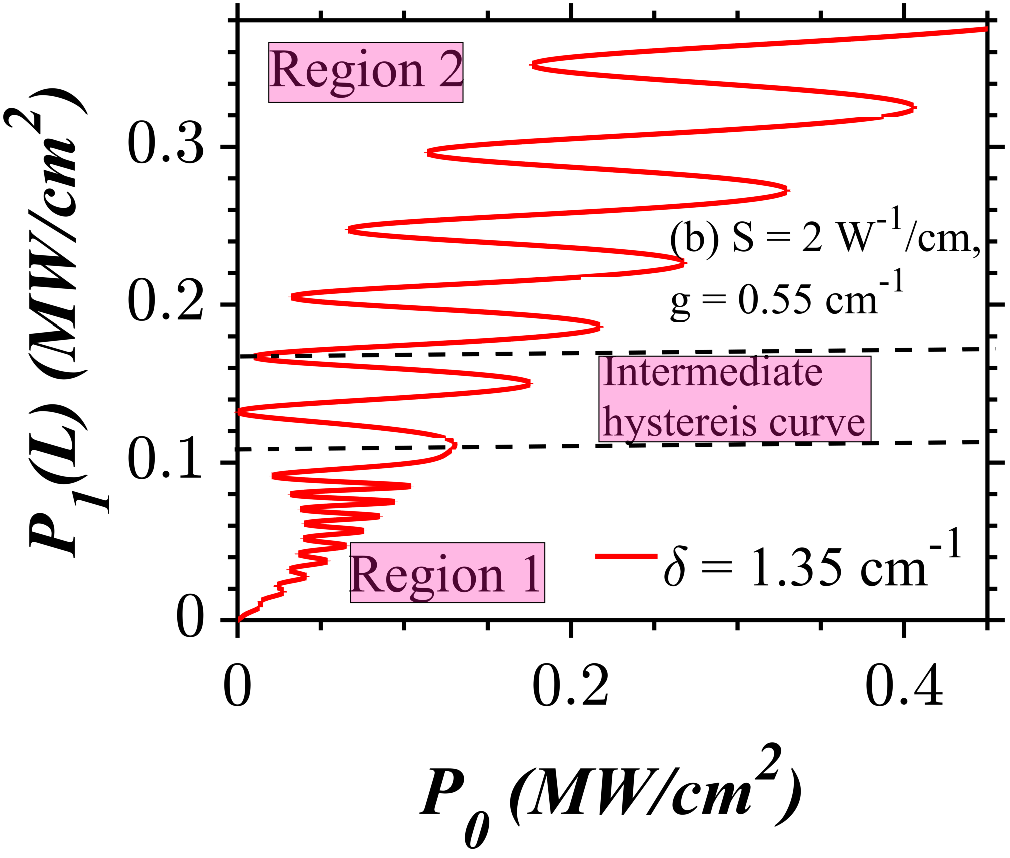}
	\caption{Formation of low-power mixed OM with vortex-like envelope in the input-output characteristics of a broken PTFBG with SNL at $L = 70$ $cm$ and $S = 2$ $W^{-1}/cm$. The direction of light incidence is right.}
	\label{fig11}
\end{figure}

\begin{figure}
	\centering	\includegraphics[width=0.5\linewidth]{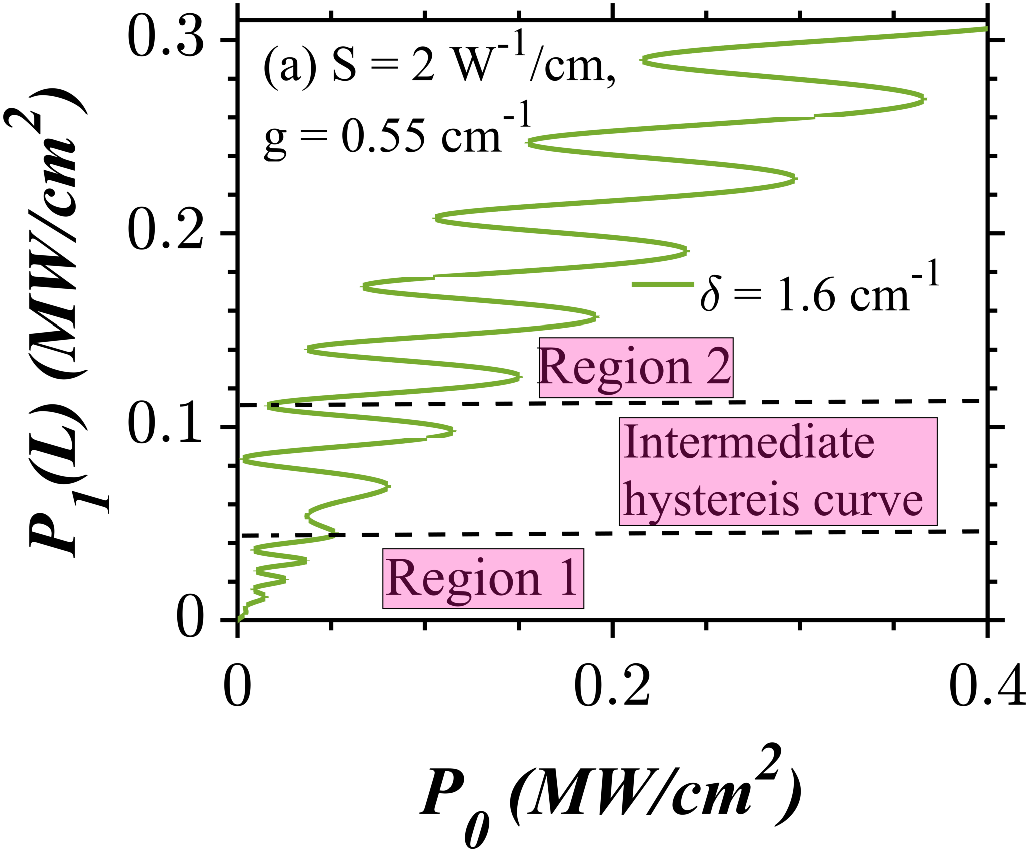}\includegraphics[width=0.5\linewidth]{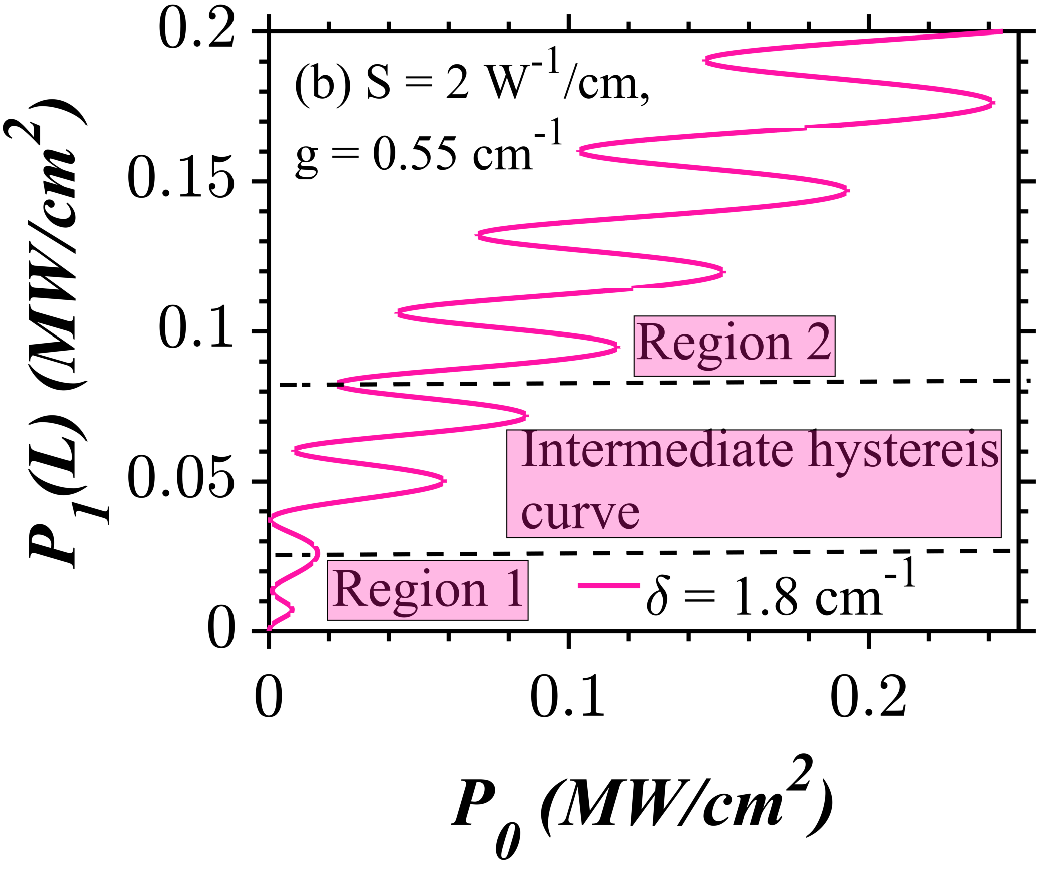}
	\caption{Input-output characteristics of a broken PTFBG with SNL at $L = 70$ $cm$ and $S = 2$ $W^{-1}/cm$. (a) and (b) Increase and decrease in the number of S-shaped and ramp-like hysteresis curves, respectively, of a low-power mixed OM curve under frequency detuning. The direction of light incidence is right. }
	\label{fig12}
\end{figure}

For $\delta < 1$ $cm^{-1}$, the system exhibits ramp-like OM curves for  the given values of input intensities, as shown in Figs. \ref{fig10}(a) and (b).  The switch-up and down intensities of the low-power ramp-like OM curve drift towards the higher and lower intensity sides, respectively, marking an unusual phenomenon.  Typically, in any OM curve, they drift toward the higher-intensity side, which is not the case with the present system. Simultaneous drift in the switch-up and switch-down intensities towards the higher and lower intensity sides leads to a vortex-like envelope, as shown in Figs. \ref{fig10}(a) and (b). The existence of OM curves with a vortex-like envelope, as observed in our study, represents an extraordinary and unprecedented finding not reported in existing literature. This remarkable phenomenon is feasible due to the intricate interplay between SNL and the right light incidence condition. When the detuning parameter is increased, the switching intensities pertaining to different stable branches of these peculiar OM curves with vortex-like envelope decrease, as shown in Fig. \ref{fig10}(b).

\subsection{Low power mixed OM curves with vortex-like envelope at $L = 70$ cm and detuning $1<\delta < 1.5$     $cm^{-1}$}
\label{Sec:4F}

When there is a significant amount of deviation of the operating wavelength from the Bragg wavelength via frequency detuning, the system begins to generate OM curves that exhibit a combination of ramp-like OM curves with a vortex-like envelope for low values of input intensities in region 1 and S-shaped OM curves for high values of input intensities in region 2, as shown in Fig. \ref{fig11}(a). An intermediate hysteresis curve with a near-zero switch-down intensity separates the ramp-like portion of a mixed OM curve in region 1 from the S-shaped OM curves in region 2. It exhibits a sharp increase in output intensities in the lower branch, contrasting with a gradual increase in the upper branch. It is noteworthy that the formation of OM curves, where one of its branches features near-zero switch-down intensities, has been reported in anti-directional couplers \cite{govindarajan2019nonlinear} and PTFBGs with inhomogeneous NL profiles \cite{sudhakar2022inhomogeneous}. However, the concurrent presence of mixed OM curves showcasing a vortex-like envelope and a hysteresis curve with near-zero switching intensities is a unique discovery not mentioned in earlier studies. 

The switch-up, and down intensities of various stable branches of a mixed OM curve decrease further with an increase in the value of the detuning parameter to $\delta = 1.35$ $cm^{-1}$, as shown in Fig. \ref{fig11}(b). Furthermore, it was observed that there is a considerable reduction in the number of ramp-like stable branches in region 1. It is also noticeable that the number of S-shaped hysteresis curves in region 2 is increasing at the same condition.
\subsection{Low power mixed OM curves at $L = 70$ cm for $\delta > 1.5$ $cm^{-1}$}
\label{Sec:4G}

The number of low-power ramp-like hysteresis curves (region 1) keeps decreasing with an increase in the detuning parameter, as shown in Figs. \ref{fig12} (a) and (b). Simultaneously, there is an increase in the number of low-power S-like hysteresis curves (in region 2). We observe only one ramp-like hysteresis curve at $\delta = 1.8$ $cm^{-1}$ in  Fig. \ref{fig12} (b).  The switching from the ramp-like hysteresis curve to the intermediate hysteresis curve and then to the S-shaped hysteresis curves takes place at an extremely low-intensity values. 

 The results of these observations indicate that under the variation in the detuning parameter, the system may exhibit two distinct varieties of low-power mixed OM curves, which may be different from one another. The first type of mixed OM curves shows a higher number of ramp-like hysteresis curves within the given range of input intensities. As compared to the first type of mixed OM curve, the second type features just a few ramp-like hysteresis curves, and there are more S-shaped hysteresis curves within the given range of input intensities.

\section{Advantages of OM over OB in memory applications}
	\label{Sec:5}
In the context of modern-day all-optical signal processing, the demand is for optical memories with high-density, high-speed, smaller dimensions, low-power consumption, higher degree of tunability, reconfigurability, and integrability to effectively store and buffer large telecommunication data \cite{hill2004fast}. In the future, integrated optical memory technologies could emerge as a compelling choice for energy-efficient and compact data storage \cite{alexoudi2020optical}. Two-level binary optical memories are commonly employed in contemporary all-optical networks for data buffering \cite{adams1987physics}. For example, optical static RAM layouts have primarily utilized bistable optical devices, and the realization of optical dynamic RAM with high speeds and low-power consumption with all-optical multistable devices still remains as a challenging one \cite{pleros2008optical}.  With the evolution of fiber optic transmission systems, the demand for processing multiple packets of information with a single device has grown. Substantial advancements have been made in the field of all-optical memories \cite{temnykh2010optical}, particularly those involving multi-level configurations \cite{tohari2020optical,zhang2017optical}, over the last decade. A recent review article on optical memory technology suggests that the domain has progressed from millimeter-long semiconductor optical amplifiers to micro-rings, microdisks, and has further advanced to III–V photonic crystal cavities \cite{alexoudi2020optical}. 

The peculiar features observed in the OM curves within the proposed system suggest that they could provide an alternative route in the continually evolving landscape of optical memories. The OM curves appearing in Fig. \ref{fig7} may not only showcase intriguing behaviors but may also provide the potential for encoding and achieving high information density in optical memory applications. This may allow for a higher data storage capacity compared to systems with bistable behavior. Also, these OM curves illustrated in Fig. \ref{fig7} represent an attractive prospect for low-power memory applications. Mixed OM curves exhibit gradual transitions between the ramp-like and S-shaped portion of the hysteresis curve. This may aid in more controlled switching between memory states which in turn may reduce the errors occurring in transmission and processing of data during read and write operations. The dynamics of ramp-like and mixed OM curves with a unusual vortex-like envelope can lead to unconventional and advanced memory functionalities. This complexity can be explored for the design of innovative all-optical memory designs in near future. It is pertinent to note that the nature of the OM curves can be easily tailored by adjusting one or more system parameters. This tunable property can be explored for implementing dynamic and re-configurable memory systems whose characteristics can be adjusted to meet specific requirements of data processing in optical computing applications.

	\section{Conclusions}
	\label{Sec:6}
	\begin{table*}[h!]
		\caption{Types of hysteresis curves admitted by the PTFBG with SNL in the broken $\mathcal{PT}$- symmetric regime.}
		\begin{center}
				\begin{tabular}{c c c c c c c c}
					\hline
					\hline
					{Nature of the}&{Classification of}&{Value of S} & {Range of $\delta$ }&{Nature of variations}&{If OM, whether} &{Light}&{Fig. no} \\
					{hysteresis}&{the curve}&{($W^{-1}/cm$)} & {($cm^{-1}$)}&{in the output} &{hysteresis width }&{incidence}&{} \\
					{curve}&{}&{}&{}&{intensities}&{increases or}&{condition}&{}\\
						{}&{}&{}&{}&{}&{ decreases? vs $P_0$}&{}&{}\\
					\hline

				{S-shaped OB}&{Type -- I}&{0.5}&{$<$ 0.32}&{gradual}&{--} &{LI}&{Figs. \ref{fig1} and \ref{fig5}(a)}\\
				{curves}&{}&{0.5 -- 0.8}&{$<$ 0.32}&{}&{}&{RI}&{Fig. \ref{fig6}}\\
				{}&{}&{}&{}&{}&{}&{}&{}\\

				{ramp-like OB}&	{Type -- II}&{1}&{$<$ 0.75}&{sharp}&{--}&{LI}&{Fig. \ref{fig2}(a)}\\
			{curves}&{}&{0.8 -- 1.1}&{$<$ 0.75}&{}&{}&{RI}&{Fig. \ref{fig7}}\\
				{}&{}&{}&{}&{}&{}&{}&{}\\
				
					{ramp-like OM}&{Type -- III}&	{0.5}&{$\le$ 1}&{sharp} &{increases}&{LI}&{Fig. \ref{fig4}(a) }\\
				{curves }&{}&{}&{}&{}&{}&{}&{}\\

				{}&{}&	{1}&{$\ge$ 0.5}&{} &{}&{LI}&{Figs. \ref{fig2}(b), \ref{fig4}(b) and \ref{fig5}(b)}\\

					{}&{}&	{2}&{$<$ 1.2}&{}&{}&{LI}&{Figs. \ref{fig3}(a), \ref{fig4}(c) and \ref{fig5}(c)}\\
					{}&{}&{1 -- 2}&{$<$ 1.2}&{}&{}&{RI}&{Fig. \ref{fig8}}\\
				{}&{}&{}&{}&{}&{}&{}&{}\\
					{}&{}&{}&{}&{}&{}&{}&{}\\

					{mixed OM}&{Type -- IV}&	{2}&{$>$ 1.2}&{sharp at}&{increases at}&{LI}&{Figs. \ref{fig3}(b), (c), \ref{fig4}(d), (e) }\\
						{curves}&{}&{}&{}&{low intensities}&{low intensities}&{}&{\ref{fig5}(d)}\\
					{ }&{}&{}&{}&{and}&{and}&{}&{}\\

				{}&{}&{1.2 -- 2}&{$>$ 1.2}&{gradual at}&{decreases at}&{RI}&{Fig. \ref{fig9} }\\
				{}&{}&{2}&{$>$ 1.5}&{high intensities}&{high intensities}&{}&{Fig. \ref{fig12}}\\
				{}&{}&{}&{}&{}&{}&{}&{}\\
				{}&{}&{}&{}&{}&{}&{}&{}\\

					{ramp-like OM }&{Type -- V}&	{2}&{$\le$ 1.2}&{sharp} &{increases}&{RI}&{Fig. \ref{fig10}}\\
				{curves with  }&{}&{}&{}&{}&{}&{}&{}\\
				{vortex-like }&{}&{}&{}&{}&{}&{}&{}\\
				{envelope}&{}&{}&{}&{}&{}&{}&{}\\
				{}&{}&{}&{}&{}&{}&{}&{}\\
				
					{mixed OM}&{Type -- VI}&	{2}&{ 1.2 -- 1.5}&{sharp at} &{increases at}&{RI}&{Figs. \ref{fig11}}\\
				{curves with  }&{}&{}&{}&{low intensities}&{low intensities}&{}&{}\\
					{vortex-like}&{}&{}&{}&{and}&{and}&{}&{}\\
				{envelope }&{}&{}&{}&{gradual at}&{decreases at}&{}&{}\\
				{}&{}&{}&{}&{high intensities}&{high intensities}&{}&{}\\
					\hline\hline
			\end{tabular}
			\label{tab2}
		\end{center}
	\end{table*}

\begin{table*}[h!]
	\caption{\centering Difference between other PTFBG configurations without SNL and present system} 
	\begin{center}
			\begin{tabular}{c c c c}
				\hline
				\hline
				{Nature of the hysteresis }&{Classification of}&{PTFBG configurations } & {present system}  \\
				{curve}&{the curve}&{without SNL} & {}  \\
				\hline
				{S-shaped OB/OM }&{Type - I}&{supported regimes: conventional }&{supported regimes: unbroken }\\
				{curves}&{}&{and unbroken regimes}&{and broken regimes}\\
				{}&{}&{}&{}\\
				{}&{}&{Novel finding: cannot occur in broken regime}&{cannot appear in conventional regime}\\
				
				{}&{}&{}&{}\\
				{ramp-like OB/OM }&{Type - II and III}&{supported regime:  }&{supported regime: conventional,  }\\
				{curves}&{ }&{broken only}&{unbroken and broken regimes}\\
				{}&{}&{}&{}\\
				{If OM, whether the  }&{Type - III}&{decreases}&{increases}\\
				{width increases or decreases vs $P_0$}&{}&{}&{}\\

				{}&{}&{}&{}\\
				{mixed OM }&{Type -- IV}&{cannot occur }&{supported $\mathcal{PT}$-symmetric regime:  }\\
				{curve}&{}&{}&{unbroken and broken regimes}\\
				{}&{}&{}&{}\\

				\hline\hline
		\end{tabular}
		\label{tab10}
	\end{center}
\end{table*}

\begin{table*}
	
	\caption{Difference between hysteresis curves in the unbroken and broken regimes, pertaining to PTFBGs with SNL. Note that $S$ can be tuned from 0.5 to 2 $W^{-1}/cm$ and the range is classified into the following ranges for convenience. Low: S$< 0.5$, moderately high: 0.5 $< S < 1$, high: 1 $<S<$ 1.5 and very high: $S > 1.5$ $W^{-1}/cm$.  Similarly, the values of detuning parameter in the range $0.5 < \delta < 0.75$ $cm^{-1}$ can be taken as moderately high values. The values below and above this range is considered to be low and high, respectively.}
	\begin{center}
			\begin{tabular}{c c c c}
				\hline
				\hline
				{Nature of the  }&{Classification of}&{unbroken} & {broken }  \\
				{hysteresis curve}&{the curve}&{ regime \cite{raja2022saturate} } & {regime}  \\
				\hline
				{S-shaped OB/OM curves}&{Type - I}&{Values of $S$: all}&{values of $S$: low}\\
				{}&{}&{Values of $\delta$: high}&{Values of $\delta$: low}\\
					{}&{}&{Range of $\delta$: narrow }&{Range of $\delta$: narrow}\\
				{}&{}&{}&{}\\
				{}&{}&{}&{}\\
				{Ramp-like OB/OM curves }&{Type - II and III}&{Values of $S$: all}&{values of $S$: moderately high}\\
				{}&{}&{only at $\delta = 0$ }&{values of $\delta$: high}\\
					{}&{}&{}&{range of $\delta$: broad}\\
				{}&{}&{}&{}\\
				
				{Mixed OM curves}&{Type -- IV}&{Values of $S$: all }&{ values of $S$: high and very high}\\
				{}&{}&{values of $\delta$: low}&{values of $\delta$: high}\\
					{}&{}&{Range of $\delta$: narrow }&{Range of $\delta$: broad}\\
				{}&{}&{}&{}\\
				{}&{}&{}&{}\\
				
				{Ramp-like OM with}&{Type -- V}&{cannot occur}&{ values of $S$: very high }\\
				{vortex-like envelope}&{}&{}&{values of $\delta$: $<1$ $cm^{-1}$}\\
				{}&{}&{}&{under right light}\\
				{}&{}&{}&{incidence condition}\\
				{}&{}&{}&{}\\
				
				{Mixed OM with}&{Type -- VI}&{cannot occur}&{ values of $S$: very high }\\
			{vortex-like envelope}&{}&{}&{values of $\delta$: 1- 1.5 $cm^{-1}$}\\
			{}&{}&{}&{under right light}\\
			{}&{}&{}&{incidence condition}\\
			{}&{}&{}&{}\\	
				
				{When $g$ increases switching intensities of}&{Type - I, II,}&{decreases}&{ increases }\\
				{ various curves increases or decreases?}&{III and IV}&{}&{}\\
				{}&{}&{}&{}\\
				
				\hline\hline
		\end{tabular}
		\label{tab12}
	\end{center}
 \end{table*}
	The broken PTFBG exhibits diverse forms of OB (OM) curves in the presence of SNL, and a summary of the findings is tabulated in Table \ref{tab2}. The variations in the nature of the hysteresis curves are achieved by independently manipulating the NL or detuning parameter. In this manner, transitions from one form of OM to another occur. The role of different system parameters in the properties of the OB and OM curves was presented earlier in Table \ref{tab9}. At this juncture, we would like to once again stress that the characteristics of the OB/OM curves admitted by the PTFBG with SNL are completely different from the rest of the PTFBG configurations, and a short summary of the differences is provided in Table \ref{tab10}. Even within the context of PTFBG with SNL, the diverse forms of OB (OM) curves admitted by the system in the broken regime are distinct from those occurring in the unbroken regime, as shown in Table \ref{tab12}. This is an additional highlight of the present article.   The formation of the standard S-shaped hysteresis curve in the broken $\mathcal{PT}$- symmetric regime, a phenomenon not reported in any other PTFBG configuration, is made feasible solely due to the presence of SNL.   An increase in the detuning parameter resulted in a significant reduction in the switching intensities. On top of this, the frequency detuning favored the transformation of ramp-like OM curves into  mixed OM curves.
	
	 The concept of right light incidence condition enabled the realization of different types of OM curves at low intensities. In this light incidence condition, the switch-up and down intensities corresponding to different stable branches of a ramp-like OM and mixed OM curves demonstrated a drift towards higher and lower intensity sides, respectively, leading to the formation of a vortex-like envelope. Such a drift stimulated a near-zero switch-down phenomenon in one of the intermediate hysteresis curves. The ramp-like (S-shaped) OM curves appeared predominantly over the S-shaped (ramp-like) hysteresis curves in a mixed OM curve for lower (higher) values of the detuning parameter.  In this fashion, the ramp-like (S-shaped) stable branches decreased (increased) under the frequency detuning in a low-power mixed OM curve. To sum up, the PTFBG with SNL offers numerous degrees of freedom to control light with light. 
	\section*{Acknowledgments}
	SVR is supported by the  Department of Science and Technology (DST)-Science and Engineering Research Board (SERB), Government of India, through a National Postdoctoral Fellowship (Grant
	No. PDF/2021/004579). AG acknowledges the support from the University Grants Commission (UGC), Government of India, for providing a Dr. D. S. Kothari Postdoctoral Fellowship (Grant
	No. F.4-2/2006 (BSR)/PH/19-20/0025). ML wishes to thank the DST-SERB for the award of a DST- SERB National Science Chair in which AG is now a Visiting Scientist (Grant No. NSC/2020/000029).

\end{document}